\documentclass[twocolumn,prx,superscriptaddress]{revtex4}
\usepackage{amsfonts}
\usepackage{amsmath}
\usepackage{amssymb}
\usepackage{graphicx}
\usepackage{bm}
\usepackage{color}
\usepackage{mathrsfs}
\usepackage{amsbsy}
\usepackage{latexsym,epsfig,graphicx}
\usepackage{dcolumn}
\usepackage{subfigure}
\usepackage{comment}
\usepackage{xspace}
\usepackage{epstopdf}
\usepackage{tabularx}
\usepackage{longtable}
\usepackage[colorlinks=true, letterpaper=true, pdfstartview=FitV, linkcolor=blue, citecolor=blue, urlcolor=blue]{hyperref}
\usepackage[normalem]{ulem}
\renewcommand{\Re}{\operatorname{Re}}
\renewcommand{\Im}{\operatorname{Im}}

\newcommand{\Tr}{\mathrm{Tr}}
\newcommand{\RNum}[1]{\uppercase\expandafter{\romannumeral #1\relax}}

\begin{document}
\title{Correspondence between winding numbers and skin modes in non-hermitian systems}

\author{Kai Zhang}\thanks{These two authors contributed equally}
\affiliation{Beijing National Laboratory for Condensed Matter Physics, and Institute of Physics, Chinese Academy of Sciences, Beijing 100190, China}
\affiliation{University of Chinese Academy of Sciences, Beijing 100049, China}

\author{Zhesen Yang}\thanks{These two authors contributed equally}
\affiliation{Beijing National Laboratory for Condensed Matter Physics, and Institute of Physics, Chinese Academy of Sciences, Beijing 100190, China}
\affiliation{University of Chinese Academy of Sciences, Beijing 100049, China}

\author{Chen Fang}
\email{cfang@iphy.ac.cn}
\affiliation{Beijing National Laboratory for Condensed Matter Physics, and Institute of Physics, Chinese Academy of Sciences, Beijing 100190, China}
\affiliation{Songshan Lake Materials Laboratory, Dongguan, Guangdong 523808, China}
\affiliation{CAS Center for Excellence in Topological Quantum Computation, Beijing 100190, China}

\begin{abstract}
We establish exact relations between the winding of ``energy'' (eigenvalue of Hamiltonian) on the complex plane as momentum traverses the Brillouin zone with periodic boundary condition, and the presence of ``skin modes'' with open boundary condition in non-hermitian systems.
We show that the nonzero winding with respect to any complex reference energy leads to the presence of skin modes, and vice versa.
We also show that both the nonzero winding and the presence of skin modes share the common physical origin that is the non-vanishing current through the system.
\end{abstract}

\maketitle

\emph{Introduction.---}Some systems that are coupled to energy or particle sources or drains, or driven by external fields can be effectively modeled Hamiltonians having non-hermitian terms~\cite{Bender_2007,moiseyev2011non,Persson2000,Volya2003,Rotter2009,Rotter_2015,Choi2010,Diehl2011,Reiter2012}. 
For example, one may add a diagonal imaginary part in a band Hamiltonian for electrons to represent the effect of finite quasiparticle lifetime~\cite{Shen2018,Zhou2018,Papaj2019,Kozii2017}.
One may also introduce an imaginary part to the dielectric constant in Maxwell equations to represent metallic conductivity in a photonic crystal~\cite{Feng2017,Mirieaar7709,Longhi2009,Zloshchastiev2016,Sounas2017,El-Ganainy2018,Bliokh2019}.
As non-hermitian operators in general have complex eigenvalues, the eigenfunctions of Schr\"odinger equations are no longer static, but decay or increase exponentially in amplitude with time~\cite{Brody2013,Gong2018}.

A topic in recent condensed-matter research is the study of topological properties in band structures, which are generally given by the wave functions, \emph{not} the energy, of all occupied bands (or more generally, a group of bands capped from above and below by finite energy gaps)~\cite{Hasan2010,Qi2011,Bernevig2013,Chiu2016,Armitage2018}.
The topological band theory has been extended to non-hermitian systems and further developed in recent years~\cite{MartinezAlvarez2018,Foa_Torres_2019,Ghatak2019,1912.10048}.
In non-hermitian systems, obviously, one immediately identifies a different type of topological numbers in bands, given by the phase winding of the ``energy'' (eigenvalue of Hamiltonian), \emph{not} the wave functions, in the Brillouin zone (BZ)~\cite{Shen2018a}.
This \emph{winding number}, together with several closely related winding numbers if other symmetries are present, give topological classification that is richer than that of their hermitian counterparts~\cite{Lee2016,Leykam2017,Gong2018,Yin2018,Ghatak2019,Xiong2018}. 
Besides winding in energy in complex plane, another unique phenomenon recently proposed in non-hermitian systems is the non-hermitian skin effect in open-boundary systems~\cite{Xiong2018,TorresPRBR,Yao2018,Yao2018a,Kunst2018,Kunst2019,Lee2019,SLonghi,1902.07217,Song2019,DengPRB,ImuraPRB,GBZZhensen,LinhuPRB,ChingHuaPRL,BergholtzPRB,SOTIPRL,KawabataPRL2019,LinhuarXiv}, which has also been verified experimentally~\cite{Brandenbourger,Xiao2019,1907.11619,1908.02759}, and a simple example of skin modes can be seen in the Supplemental Material Sec.~\ref{apdnx:1}. A typical spectrum of open hermitian system consists of a large number of bulk states, and, if at all, a small number of edge states, and as the system increases in size $L$, the numbers of the bulk and of the edge states increase as $L^d$ and $L^{d-n}$ respectively, where $d$ is the dimension and $0<n\le{d}$. 
However, in certain non-hermitian systems, a finite fraction, if not all, of eigenstates are concentrated on one of the edges.
These non-hermitian skin modes decay exponentially away from the edges just like edge states, but their number scales as the volume ($L^d$), rather than the area, of the system~\cite{skinmodes_note}.

\begin{figure}
\begin{centering}
\includegraphics[width=1\linewidth]{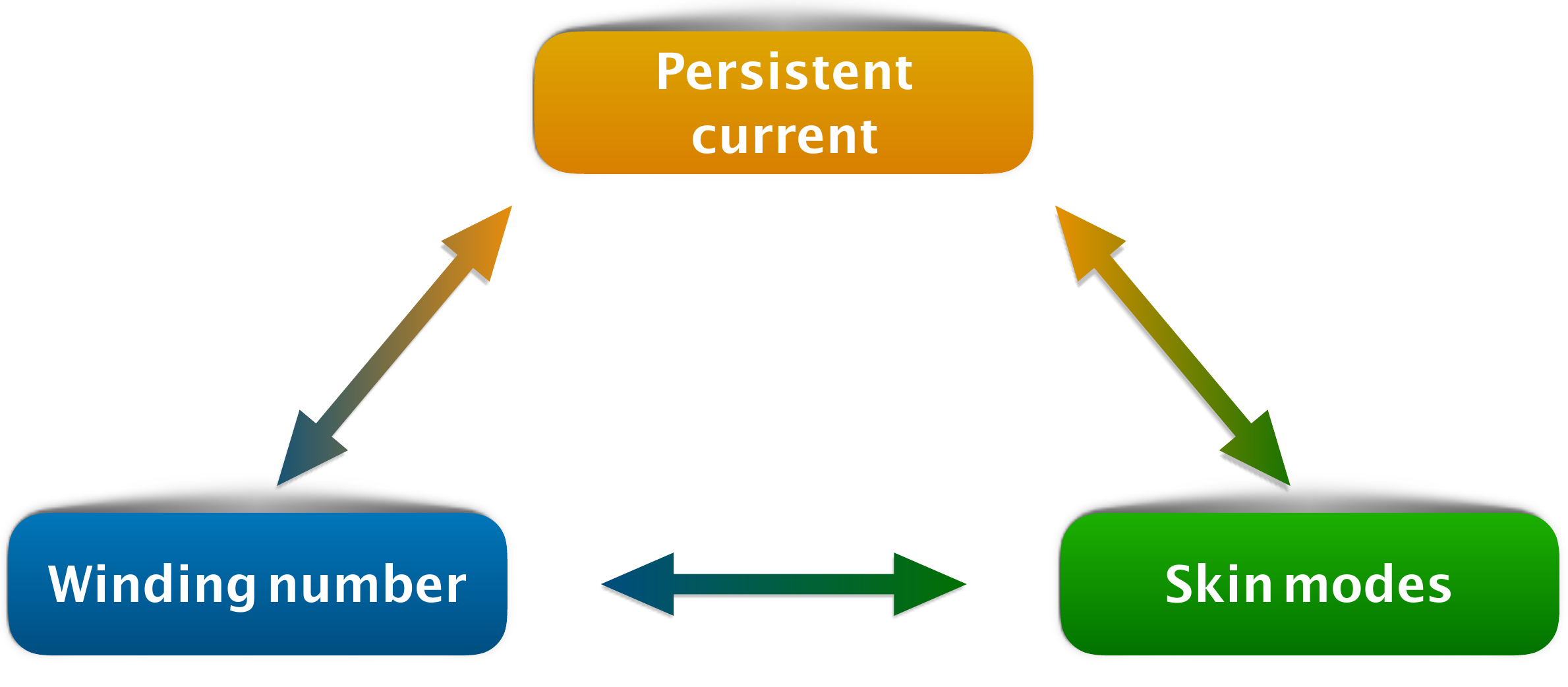}
\par\end{centering}
\protect\caption{\label{fig:1}The reciprocal relations among the three phenomena unique to non-hermitian systems: the non-vanishing persistent current, nonzero winding number of energy and the presence of skin modes. 
	The validity of any one is the sufficient and necessary condition for the validity of the other two.}
\end{figure}

In this Letter, we show an exact relation between the new quantum number, \textit{i.~e.}, the winding number of energy with periodic boundary, and the existence of skin modes with open boundary, for any one-band model in one dimension.
To do this, we first extend the one-band Hamiltonian with finite-range hopping $H(k)$ to a holomorphic function $H(z)=P_{n+m}(z)/z^m$ ($n,m>0$)~\cite{holo_note}, where $P_{n+m}(z)$ is a $(n+m)$-polynomial, and the Brillouin zone maps to unit circle $|z|=1$ (or $z=e^{ik}$). 
The image of the unit circle under $H(z)$ is the spectrum of the system with periodic boundary, and generally, it forms a loop on the complex plane, $\mathcal{L}_{\mathrm{BZ}}\in\mathbb{C}$.
Then we show that as long as $\mathcal{L}_{\mathrm{BZ}}$ has finite interior, or roughly speaking encloses finite area, skin modes appear as eigenstates with open boundary condition; but when $\mathcal{L}_{\mathrm{BZ}}$ collapses into a curve having no interior on the complex plane, the skin modes disappear.
In other words, skin modes with open boundary appear if and only if there be $E_b\in\mathbb{C}$ with respect to which $\mathcal{L}_{\mathrm{BZ}}$ has nonzero winding.
Finally, we show that the winding of the periodic boundary spectrum, and hence the presence of skin modes with open boundary, are related to the total persistent current of the system.
We prove that if the current vanishes for all possible state distribution functions $n(H,H^\ast)$, the winding and the skin modes also vanish, and vice versa.
The relations we establish among nonzero winding, presence of skin modes and non-vanishing current are summarized in Fig.~\ref{fig:1}.
Some of the results are extended to 1D models with multiple bands.

\begin{figure}
\begin{centering}
\includegraphics[width=1\linewidth]{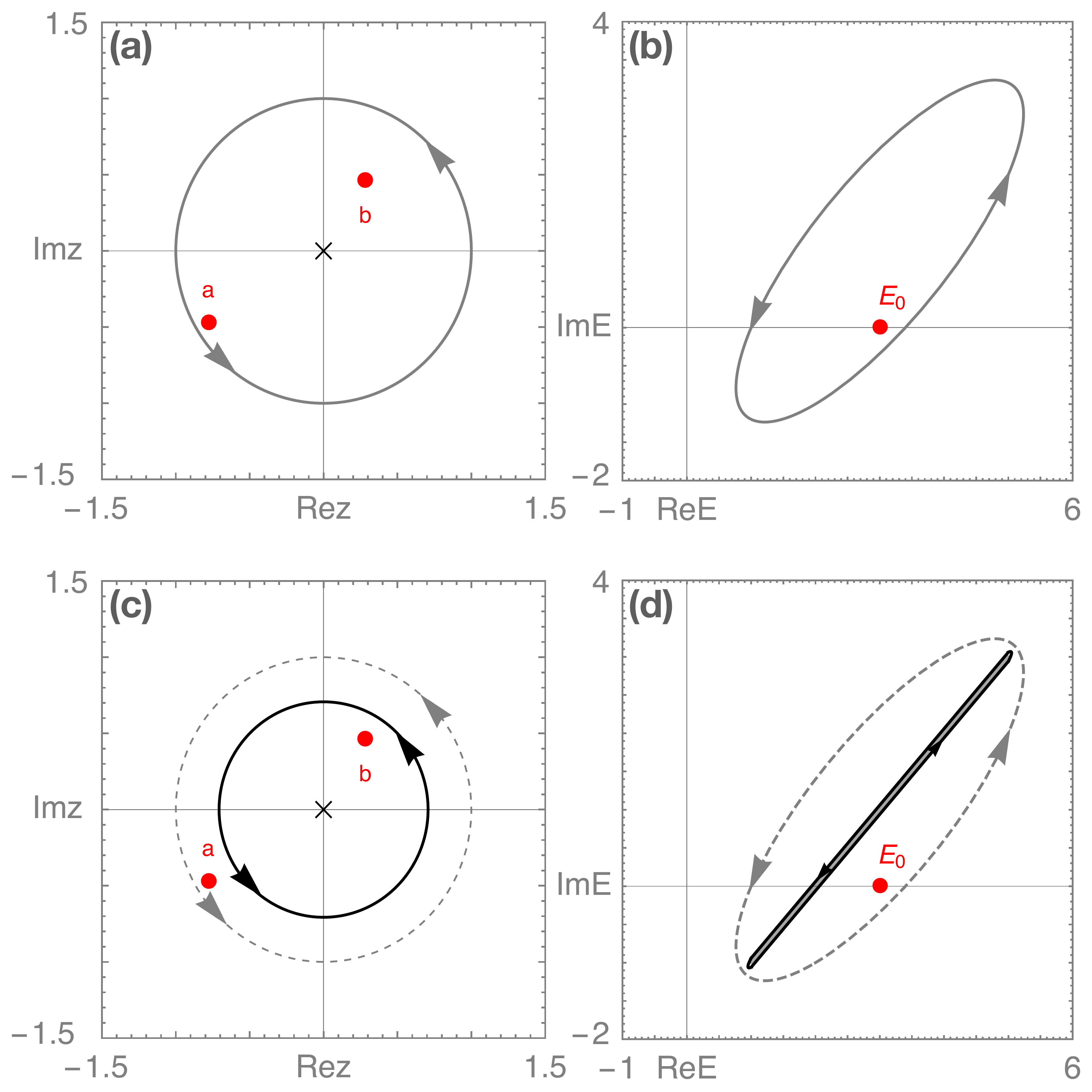}
\par\end{centering}
\protect\caption{\label{fig:2}We show the BZ (a) with periodic-boundary spectrum (b), and GBZ (c) with open boundary spectrum (d) for the model $H(z)=(2i z^2+(3+i) z+1)/z$, and the red dot $E_0=H(z=a)=H(z=b)=3$ is the reference energy with respect to which winding is defined. In (a)(c) the red dots represent the zeros of $H(z)-E_0=0$, and the cross denotes the pole. We remark that the orientation of GBZ in (c) is arbitrarily chosen.}
\end{figure}

\emph{Hamiltonian as holomorphic function.---}We start with an arbitrary one-band tight-binding Hamiltonian in one dimension, only requiring that hoppings between $i$ and $j$-sites only exist within a finite range $-m\le{}i-j\le{}n$.
\begin{equation}
H=\sum_{i,j}t_{i-j}|i\rangle\langle{j}|=\sum_{k\in\mathrm{BZ}}H(k)|k\rangle\langle{k}|,
\end{equation}
where $H(k)=\sum_{r=-m,\dots,n}t_r(e^{ik})^r$ is the Fourier transformed $t_{r}$ ($t_0$ being understood as the onsite potential).
For periodic boundary condition, we have $0\le{k}<2\pi$, and $e^{ik}$ moves along the unit circle on the complex plane.
For future purposes, we define $z:=e^{ik}$, and consider $z$ as a general point on the complex plane.
Therefore for each Hamiltonian $H(k)$, we now have a holomorphic function
\begin{equation}\label{eq:ZHam}
H(z)=t_{-m}z^{-m}+\dots+t_nz^n=\frac{P_{m+n}(z)}{z^m},
\end{equation}
where $P_{m+n}(z)$ is a polynomial of order $m+n$.
$H(z)$ has one composite pole at $z=0$, the order of which is $m$, and has $m+n$ zeros, \textit{i.~e.}, the zeros of the $(m+n)$-polynomial.
Along any oriented loop $\mathcal{C}$ and any given reference point $E_b\in\mathbb{C}$, one can define the winding number of $H(z)$
\begin{equation}\label{eq:winding}
w_{\mathcal{C},E_b}:=\frac{1}{2\pi}\oint_{\mathcal{C}}\frac{d}{dz}\arg[H(z)-E_b]dz.
\end{equation}
Specially, for $\mathcal{C}=\mathrm{BZ}$, $w_{\mathcal{C},E_b}$ is the winding of the phase of $H(z)-E_b$ along BZ, considered as a new topological number unique to non-hermitian systems~\cite{Lee2016,Leykam2017,Shen2018a,Gong2018,Yin2018,Xiong2018,KawabataPRX,Ghatak2019}.
The Cauchy principle relates the winding number of any complex function $f(z)$ to the total number of zeros and poles enclosed in $\mathcal{C}$, that is,
\begin{equation}\label{eq:4}
w_{\mathcal{C},E_b}=N_{zeros}-N_{poles},
\end{equation}
where $N_{zeros,poles}$ is the counting of zeros (poles) weighted by respective orders. See Fig.~\ref{fig:2}(a,b) for the pole, the zeros and the winding of $\mathcal{L}_{\mathrm{BZ}}$ for a specific Hamiltonian.
In fact, we always have $N_{poles}=m$, so that the winding number is determined by the number of zeros of $P_{m+n}(z)-z^mE_b$ that lie within the unit circle.
As we will see later, the advantage of extending the Hamiltonian into a holomorphic function lies in exactly this relation between the winding numbers and the zeros.

\emph{Generalized Brillouin zone.---}In Ref.~\cite{Xiong2018,Yao2018,Lee2019}, it is shown that energy spectrum of certain non-hermitian systems with open boundary may deviate drastically from that with periodic boundary, due to the presence of skin modes~\cite{Yao2018,Yao2018a,Kunst2018}.
Furthermore, in Ref.~\cite{Yao2018,Yokomizo2019}, the authors introduce a new concept of generalized Brillouin zone to signify the difference between periodic and open boundary: instead of evaluating $H(z)$ along BZ, the open-boundary energy spectrum is recovered as one evaluates $H(z)$ on another closed loop called GBZ as $L$ goes to infinity.
The GBZ is determined by the equation
\begin{equation}\label{eq:GBZ}
\mathrm{GBZ}:=\{z||H^{-1}_m(H(z))|=|H^{-1}_{m+1}(H(z))|\},
\end{equation}
where $H_i^{-1}(E)$'s satisfying $|H^{-1}_i(E)|\le|{H}^{-1}_{i+1}(E)|$ are the $m+n$ branches of the inverse function of $H(z)$.
(In Ref.~\cite{Yokomizo2019}, $m=n$ is assumed, and we extend the results to $m\neq{n}$ cases in the Supplemental Material Sec.~\ref{apdnx:2}.)
We emphasize that using GBZ, one can compute the open boundary spectrum of systems of large or infinite size by solving some algebraic equations such as Eq.(\ref{eq:GBZ}), a process we sketch using the following steps.
To begin with, one finds the inverse functions of $H(z)$, and orders them in ascending amplitude,  thus obtaining $H^{-1}_i(E)$, where $i=1,...,m+n$ because the $P_{m+n}(z)-Ez^m$ is an order $m+n$-polynomial of $z$.
Then, as there are two variables $(\Re(E),\Im(E))$ in Eq.(\ref{eq:GBZ}), by codimension counting its solution on the complex plane forms one or several close loops, which are nothing but the open boundary energy spectrum.
Finally, one substitutes these solutions back into $H^{-1}_m(E)$.
It is noted that if we are only interested in the spectrum, we may stop at the second last step, but we need GBZ in order to articulate some of our key results.

With GBZ thus defined, we state our central result (for proof see the Supplemental Material Sec.~\ref{apdnx:3}): GBZ is the closed curve in complex plane that encloses the pole (at the origin) of order $m$ and exactly $m$ zeros of $P_{m+n}(z)-Ez^m$ for arbitrary $E\in \mathbb{C}$~\cite{GBZ_note}.
This seemingly technical result has following consequences.
First, this means within GBZ the total number of zeros and poles (weighted by respective orders) cancel, so that the winding of $H(z)-E$ vanishes.
Next, the arbitrariness of $E$ ensures that GBZ is invariant under a shift of energy origin in the complex plane $H(z)\rightarrow{H}_z-E_b$.
Combining these two points, we see that the image of GBZ under $H(z)$ on the complex plane, denoted by $\mathcal{L}_{\mathrm{GBZ}}$, has zero winding with respect to any $E_b\in\mathbb{C}$, or symbolically,
\begin{equation}\label{eq:w=0}
w_{\mathrm{GBZ},E_b}=0,
\end{equation}
where the orientation of GBZ is defined in the Supplemental Material Sec.~\ref{apdnx:3}. Therefore, we finally see that the open-boundary spectrum of $H(z)$ cannot be a circle or eclipse like the periodic-boundary counterpart, and it cannot even form a loop enclosing any finite area, because in that case one can choose $E_b$ inside that area so that the winding of $\mathcal{L}_{GBZ}$ with respect to $E_b$ is nonzero.
The only possibility is that $\mathcal{L}_{\mathrm{GBZ}}$ \emph{collapses} into an arc as shown in Fig.~\ref{fig:2}(d).
In this specific example ($m=n=1$ and see caption for parameters), we plot the GBZ in Fig.~\ref{fig:2}(c) and $\mathcal{L}_{\mathrm{GBZ}}$ in Fig.~\ref{fig:2}(d) as $z$ moves counterclockwise along GBZ.
We see that while GBZ is more or less a circle, its image $\mathcal{L}_{\mathrm{GBZ}}$ keeps ``back-stepping'' itself: except for a few turning and branching points, any point in $\mathcal{L}_{\mathrm{GBZ}}$ has two or an even number of pre-images in the GBZ, so that the end result looks like more connected segments of curves than a closed loop.

\begin{figure*}
\begin{centering}
\includegraphics[width=1\linewidth]{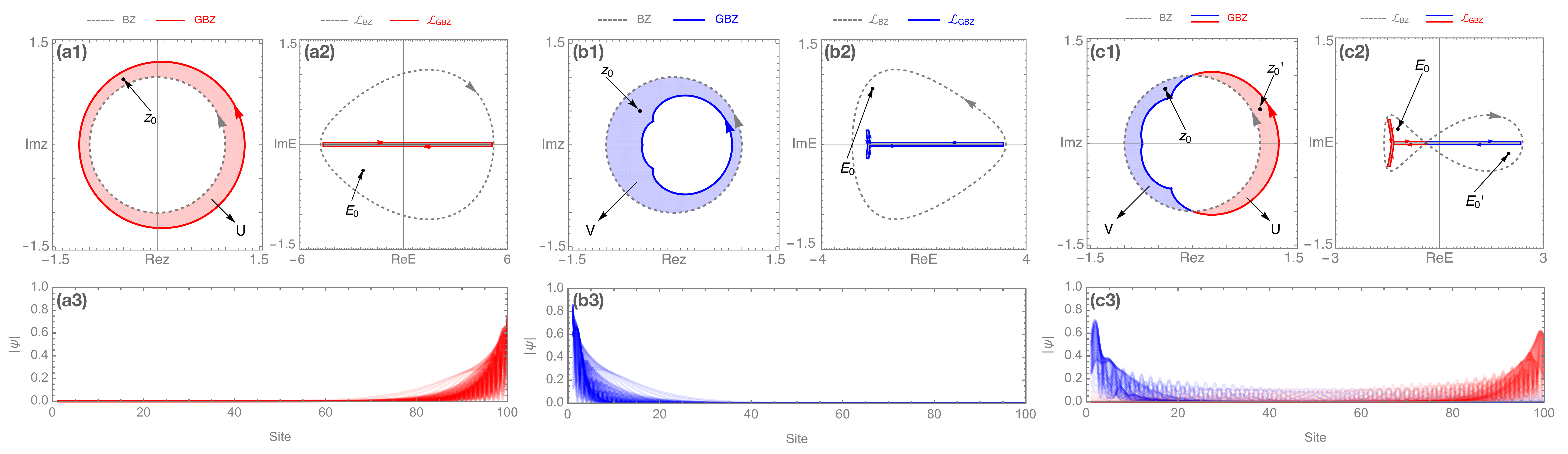}
\par\end{centering}
\protect\caption{\label{fig:3}BZ and GBZ, periodic- and open-boundary spectra, and all normalized eigenfunctions for open boundary are plotted for $H(z)=z^{-2}/5+3z^{-1}+2z$ in (a1-a3), $H(z)=z^{-2}/5+z^{-1}+2z$ in (b1-b3) and $H(z)=2z^{-2}/5+z^{-1}+z$ in (c1-c3). 
	The regions inside BZ (GBZ) and outside GBZ (BZ) are colored in blue (red), and the eigenfunctions corresponding to points on GBZ outside (inside) BZ are plotted as red (blue) curves. 
	$z_0,z_0'$ are randomly chosen points in the red and the blue regions, respectively, and $E_0=H(z_0), E'_0=H(z'_0)$.}
\end{figure*}

\emph{Skin modes and nonzero winding numbers.---}
GBZ not only gives the open boundary spectrum, but also yields information on the eigenstates with open boundary~\cite{Yao2018,Yokomizo2019}.
In fact, each point $z\in{\mathrm{GBZ}}$ represents an eigenstate, the wave function of which is in the form $\langle{s}|\psi(z)\rangle\propto{|z|^s}$, where $s=1,\dots,L$ labels the sites.
When $|z|>1$ ($|z|<1$), the wave function is concentrated near $(s=1)$-edge ($(s=L)$-edge) and exponentially decays with distance from the edge [see Fig.~\ref{fig:3}(a3, b3, c3) for examples].
Therefore, any part of GBZ that lies within (without) the unit circle corresponds to a set of skin modes.
In extreme cases, when the entire GBZ is inside (outside) the unit circle, all eigenstates are skin modes on the left (right) side of the chain.
In short, any deviation of GBZ from BZ signifies the existence of skin modes.

For a given $H(z)$, if $w_{\mathrm{BZ},E_b}\neq0$, then from Eq.(\ref{eq:w=0}) we have $w_{\mathrm{GBZ},E_b}=0$, hence GBZ must deviate from the unit circle, that is, skin modes must exist with open boundary.
Let us now try to prove the inverse statement: if GBZ and BZ differ from each other, then one can always find a $E_b\in\mathbb{C}$ such that $w_{\mathrm{BZ},E_b}\neq0$.
GBZ and BZ may differ from each other in three typical ways:
(i) as in Fig.~\ref{fig:3}(a1), GBZ contains the unit circle, and we define $U$ as the region inside GBZ but outside BZ (colored in red);
(ii) as in Fig.~\ref{fig:3}(b1), GBZ is contained in the unit circle, and we define $V$ as the region outside GBZ but inside BZ (colored in blue);
(iii) as in Fig.~\ref{fig:3}(c1), one part of GBZ is outside and another part inside the unit circle.
For case-(i), pick $z_0\in{U}$ and $E_0=H(z_0)$.
$z_0$ is then a zero of $H(z_0)-E_0$, and from Eq.(\ref{eq:w=0}), we know there are exactly $m$ zeros inside GBZ, so inside BZ there are at most $m-1$ zeros, and from Eq.(\ref{eq:4}) we have $w_{\mathrm{BZ},E_0}<-1\neq0$ [see example in Fig.~\ref{fig:3}(a2)].
For case-(ii), pick $z'_0\in{V}$ and $E'_0=H(z'_0)$, then use similar arguments to see $w_{\mathrm{BZ},E'_0}>1\neq0$ [see example in Fig.~\ref{fig:3}(b2)].
We postpone the proof for case-(iii) to the Supplemental Material Sec.~\ref{apdnx:4}, but mention here that for $z_0\in{U}$ and $z'_0\in{V}$, the periodic-boundary spectrum $\mathcal{L}_{\mathrm{BZ}}$, taking the shape of a fish [see Fig.~\ref{fig:3}(c2)], has opposite windings with respect to $E_0$ and $E'_0$.

\emph{Winding numbers, skin modes and persistent current.---}From the above results, we see that if and only if $\mathcal{L}_\mathrm{BZ}$ does not enclose any $E_b\in\mathbb{C}$, then the skin modes do not exist.
When this is the case, $\mathcal{L}_{\mathrm{BZ}}$ always ``back-steps'' itself just like $\mathcal{L}_{\mathrm{GBZ}}$, or more precisely, along $\mathcal{L}_{\mathrm{BZ}}$, for any small segment $\delta{H}$ centered at some $E$, there must be another segment $-\delta{H}$ centered at exactly the same $E$.
What is the physical meaning of this condition?
We show that this is equivalent to the absence of total persistent current with periodic boundary.
To define the current, we assume that the particles have some charge (taken to be unity), so the total persistent current can be derived as $J=\sum_kn_kv_k=\sum_kn_k{H'}(k)dk$, where $n_k$ is some distribution function~\cite{distribution_note}. Now we make a general physical assumption that $n_k$ only depends on the ``energy'' of the state, that is $n_k=n(H(k),H^\ast(k))$, but does not depend on $k$ explicitly. 
(Here $n$ depends on both the real and the imaginary parts of $H(k)$, so is unnecessarily holomorphic.) 
For example, the Bose distribution  $n_k=(e^{\Re[H(k)]/k_BT}-1)^{-1}$ satisfies such a condition.
When the curve $\mathcal{L}_{\mathrm{\mathrm{BZ}}}$ has no interior, we have
\begin{equation}\label{skincurrent}
J=\int_0^{2\pi}n(H,H^\ast)\frac{dH(k)}{dk}dk=\oint_{\mathcal{L}_{\mathrm{BZ}}}n(H,H^\ast)dH=0,
\end{equation}
that is, the total persistent current vanishes.
In the Supplemental Material Sec.~\ref{apdnx:5}, we prove the inverse statement that if there is any $E_b\in\mathbb{C}$ with respect to which $H(z)$ has nonzero winding, then one can always find some $n(H,H^\ast)\neq 0$ such that $J\neq 0$. 
This equivalence is intuitively understood: if a persistent current is going around a ring, then as one cuts open the ring, the charge starts concentrating on one end of the open chain. This persistent current is a linear response and vanish for any Hermitian system, which is proved in the Supplemental Material Sec.~\ref{apdnx:6}. 

\emph{Discussion and conclusion.---}So far we have established the reciprocal relations shown in Fig.~\ref{fig:1} for one-band model in one dimension.
Some of the results may be extended to the cases of more bands and/or higher dimensions.
For example, in $d$-dimension, one should consider a multi-variable holomorphic function $H(z_1,z_2,\dots,z_d): \mathbb{C}^d\rightarrow\mathbb{C}$, where $z_j:=e^{ik_j}$, and the spectrum of $H(z_1,\dots,z_d)$ is in general a continuum on the complex plane.
Are there skin modes when we have open boundary along $0<l\le{d}$ directions, but periodic boundary along the other $d-l$ directions?
We have two conjectures for two extreme cases:
(i) if $l=d$, that is, if all directions are open, skin modes vanish if and only if each component persistent current vanishes for arbitrary $n(H,H^\ast)$; and
(ii) if $l=1$, that is, if only one direction is open, the skin modes vanish if and only if the entire spectrum of $H(z_1,\dots,z_d)$ collapse into a curve having no interior.
The ``only if'' part of (i) and the ``if'' part of (ii) are only obvious, but the other parts seem not quite so.

Extension of the relation between the persistent current and the winding numbers in periodic boundary to multiple-band systems is straightforward. 
Now $H_{ab}(z)$ becomes a matrix function of $z:=e^{ik}$, where $a,b=1,\dots,n$ label the orbitals. 
The persistent current in this case becomes $J=\Tr (\hat{\rho} \hat{J})=\sum_{i=1,\dots,n}J_i$, where
\begin{equation}\label{eq:8}
J_i = \int_0^{2\pi}n(E_{i},E_{i}^*)\frac{dE_{i,k}}{dk}=\oint_{\mathcal{L}_{i,\mathrm{BZ}}}n(E_i,E^\ast_i)dE_i. 
\end{equation}
the operators $\hat{\rho}$ and $\hat{J}$ are steady-state density matrix operator and current operator, expressed as, respectively, $ \hat{\rho}=\sum_{i,k}n(E_{i,k},E_{i,k}^*)|i_k^R\rangle \langle i_k^L|$ and $ \hat{J}=\sum_{k,a,b}dH_{ab}(k)/dk |a_k\rangle \langle b_k| $. More details of derivation can be found in the Supplemental Material Sec.~\ref{apdnx:6}. While $J_i=0$ implies $J=0$, $J=0$ does \emph{not} necessitate $J_i=0$ for each $i$. 
In fact, one part of the trajectory of $E_i(k)$ may be back-stepped by another part of the trajectory of $E_{j\neq{i}}(k)$ so that their contribution to $J$ cancel out.
Therefore, $J=0$ is equivalent to the collapse of the spectrum, not of each individual band, but of all bands, into a curve that has no interior.
In more precise terms, $J=0$ for arbitrary $n(E,E^\ast)$ if and only if for any $E_b\in\mathbb{C}$ and $E_b\notin\mathcal{L}_{i,\mathrm{BZ}}$, the total winding number of all bands with respect to $E_b$ vanishes, or symbolically
\begin{equation}
\frac{1}{2\pi i}\int_0^{2\pi}\frac{d\log\det[H(z)-E_bI_{n\times{n}}]}{dk}dk=0.
\end{equation}
When there are additional conserved charges in the Hamiltonian, for example some spin component, we can simply replace the total current $J$ with the component current for each conserved charge $J_c$. 
At this point, we do not know exactly how the nonzero persistent current or the winding numbers are related to the skin modes in multi-band systems, but from physical intuition, we conjecture that $J\neq0$ implies skin modes with open boundary, and vice versa. 

In summary, we theoretically demonstrate that a one-dimensional non-hermitian Hamiltonian with open boundary condition has non-hermitian skin effect as long as the complex energy spectrum of the same Hamiltonian under periodic boundary condition makes a loop having nonzero area in the complex plane. The vanishing non-hermitian skin effect is also related to the vanishing persistent current for an arbitrary density matrix of a steady state. 

\emph{Acknowledgments.---}The work is supported by the Ministry of Science and Technology of China (Grant No.~2016YFA0302400), National Science Foundation of China (Grant No.~11674370) and Chinese Academy of Sciences (Grant No.~XXH13506-202, XDB33000000). This work is also supported by the Ministry of Science and Technology of China 973 program (Grant No.~2017YFA0303100), National Science Foundation of China (Grant No.~NSFC-11888101), and the Strategic Priority Research Program of CAS (Grant No.~XDB28000000).

\emph{Note added}.---After the initial posting of this paper to arXiv, the authors became aware of a closely related work~\cite{Okuma2019}. 

\bibliography{NonHermitianBib}

\begin{thebibliography}{65}
\bibitem [{ski()}]{skinmodes_note}%
  \BibitemOpen
  \href@noop {} {\bibinfo  {journal} {In Ref.~\cite{TorresPRBR}, the presence
  of skin modes is related to the vicinity of the system to a high-order
  exceptional point, at which the number of states collapses. Our model in
  Eq.(\ref{eq:ZHam}) becomes exceptional when $t_i=0$ for all $i<0$, and here
  we assume that our system is away from this point in parameter space}\
  }\BibitemShut {NoStop}%
\bibitem [{hol()}]{holo_note}%
  \BibitemOpen
\bibfield  {journal} {  }\href@noop {} {\bibinfo  {journal} {Holomorphic
  functions are a special case of complex function $ f(z,z^*) $, where the
  $z^\ast$ dependence is absent, or, $df(z,z^\ast)/d{z^\ast}=0$. A holomorphic
  function can hence be denoted as $f(z)$}\ }\BibitemShut {NoStop}%
\bibitem [{dis()}]{distribution_note}%
  \BibitemOpen
\bibfield  {journal} {  }\href@noop {} {\bibinfo  {journal} {We are familiar
  with distribution functions such as Boltzmann distribution, Bose
  distribution, Fermi distribution etc., but in a steady-state system, it can
  actually be any real function of energy. For example, in systems for laser
  emission the electrons have ``population inversion'' in their distribution}\
  }\BibitemShut {NoStop}%
\end{thebibliography}
\bibliographystyle{apsrev4-1}
\onecolumngrid
\newpage
\renewcommand{\theequation}{S\arabic{equation}}
\renewcommand{\thefigure}{S\arabic{figure}}
\renewcommand{\thetable}{S\arabic{table}}
\setcounter{equation}{0}
\setcounter{figure}{0}
\setcounter{table}{0}

\begin{center}
    {\bf \large Supplemental Material for ``Correspondence between winding numbers and skin modes in non-hermitian systems" }
\end{center}


\section{A simple example of skin modes}\label{apdnx:1}
\begin{figure}[htb]
	\begin{centering}
		\includegraphics[width=0.7\linewidth]{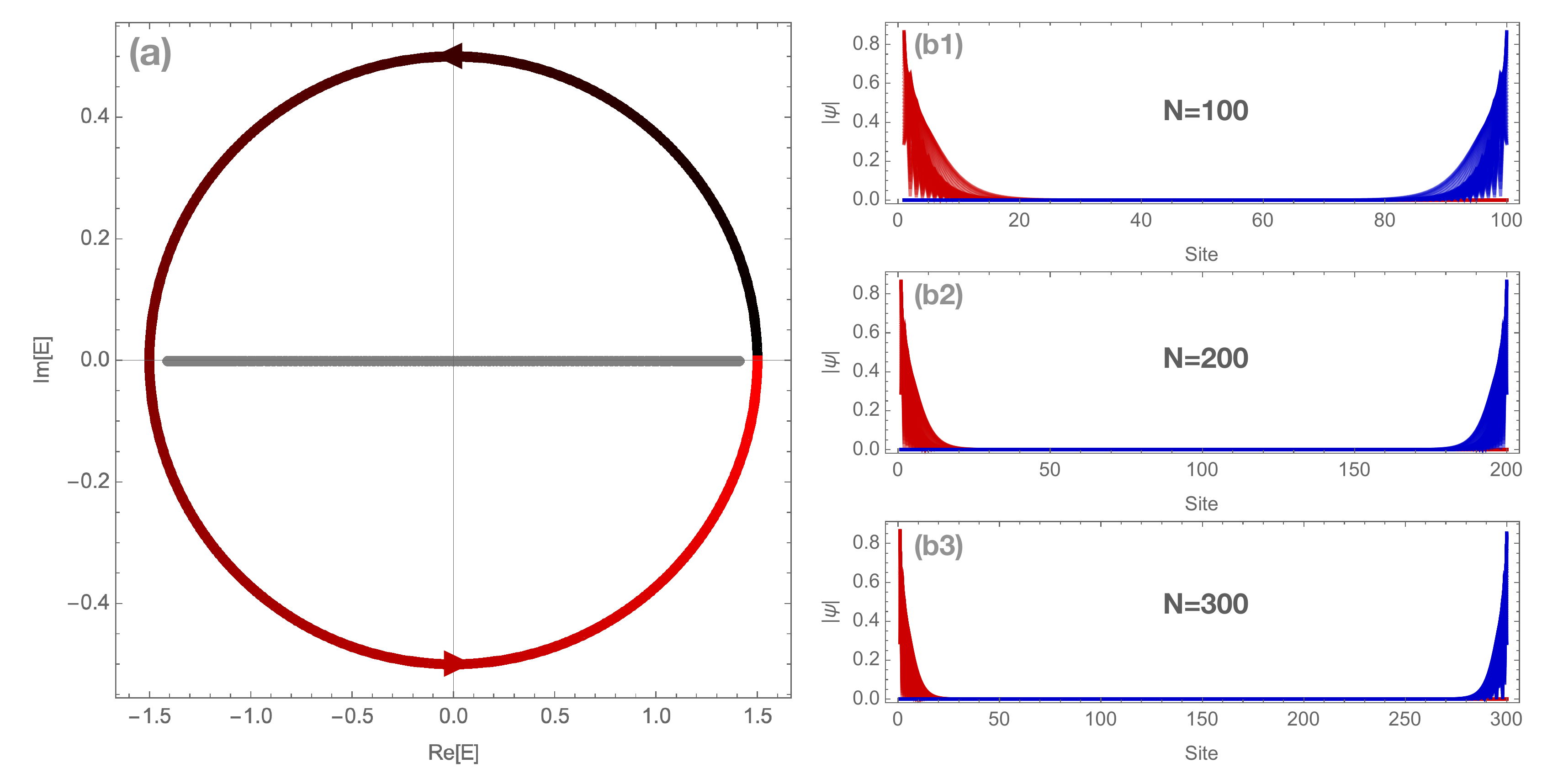}
		\par\end{centering}
	\protect\caption{\label{figs1}Spectrum and distribution of eigenvectors of Eq.~(\ref{RealSMs:I}). (a) shows the open-boundary spectrum with gray color, and periodic-boundary spectrum with the color that changes from black to red as momentum varies from $ 0 $ to $ 2\pi $. (b1), (b2), (b3) depict the distribution of all the eigenvectors on one-dimensional chains with $ N=100,200,300 $, respectively, where right eigenvectors are marked in red color, left eigenvectors in blue color. 
	}
\end{figure}

In this section, we use an example to show the non-hermitian skin modes scale as the system size $O(L)$, which is in contrast to the topological boundary states in the Hermitian systems. Consider the following non-hermitian tight-binding model in real space
\begin{equation}\label{RealSMs:I}
H = \sum_{i=1}^N \hat{c}^{\dagger}_i  \hat{c}_{i+1}+\frac{1}{2}\hat{c}^{\dagger}_{i+1}  \hat{c}_{i}, 
\end{equation}
which is a single-band model with only one internal degree per unit cell, and $ N $ is the size of the chain. The left-hopping parameter is chosen as $ 1 $, which is larger than the right-hopping parameter $ 1/2 $. Obviously, the open-boundary Hamiltonian is non-defective for arbitrary $N$. The biorthogonal eigen equations of $H$ can be  expressed as follows,
\begin{equation}
	\begin{split}
	&H|\Psi_i^r\rangle=E_i|\Psi_i^r\rangle; \\
	&H^\dag|\Psi_i^l\rangle=E_i^*|\Psi_i^l\rangle,
	\end{split}
\end{equation}
where $|\Psi_i^{r}\rangle,|\Psi_i^{l}\rangle$ represent the $i$th right and left eigenvectors, respectively. All the $N$ right eigenvectors with blue color in Fig.~\ref{figs1}(b), i.e., $|\Psi_i^r\rangle$ with $i=1,...,N$, are localized on the right edge, while all the $N$ left eigenvectors with red color in Fig.~\ref{figs1}(b), i.e., $|\Psi_i^l\rangle$ with $i=1,...,N$, are localized on the left edge of the chain. These localized eigenvectors are called skin modes. In this system, the number of skin modes is proportional to the system size. 

When the periodic boundary conditions are taken for Eq.(\ref{RealSMs:I}), the Hamiltonian can be written in the momentum space by the Fourier transformation
\begin{equation}\label{MomentumSMs:I}
H(k) = e^{i k}+e^{-i k}/2
\end{equation}
with momentum $ k $ varying from $ 0 $ to $ 2\pi $. The winding number of the spectrum can be defined as $k$ change from $0$ to $2\pi$ for arbitrarily chosen base energy, as illustrated in Fig.~\ref{figs1}(a). It seems that the winding number of $ H(k) $ is related to the presence of skin modes of $ H $. Indeed, in the main text, we strictly establish the relation between the winding of periodic-boundary spectrum and the presence of skin modes with open boundary condition. Equivalently, the presence of skin modes corresponds to the existence of nonzero area enclosed by the periodic-boundary spectrum, which is intrinsic to non-hermitian systems. 

\section{Derivation of Eq.(5)}\label{apdnx:2}

In this section, we first calculate the generalized Brillouin zone \cite{Yao2018_SM} of a heuristic single-band model and further give the general formal proof of Eq.(5) in the main text. Finally, we generalize the conclusions to the two-band model with sublattice symmetry. 

\subsection{Model}

\begin{figure}[t]
	\begin{centering}
	\includegraphics[width=0.7\linewidth]{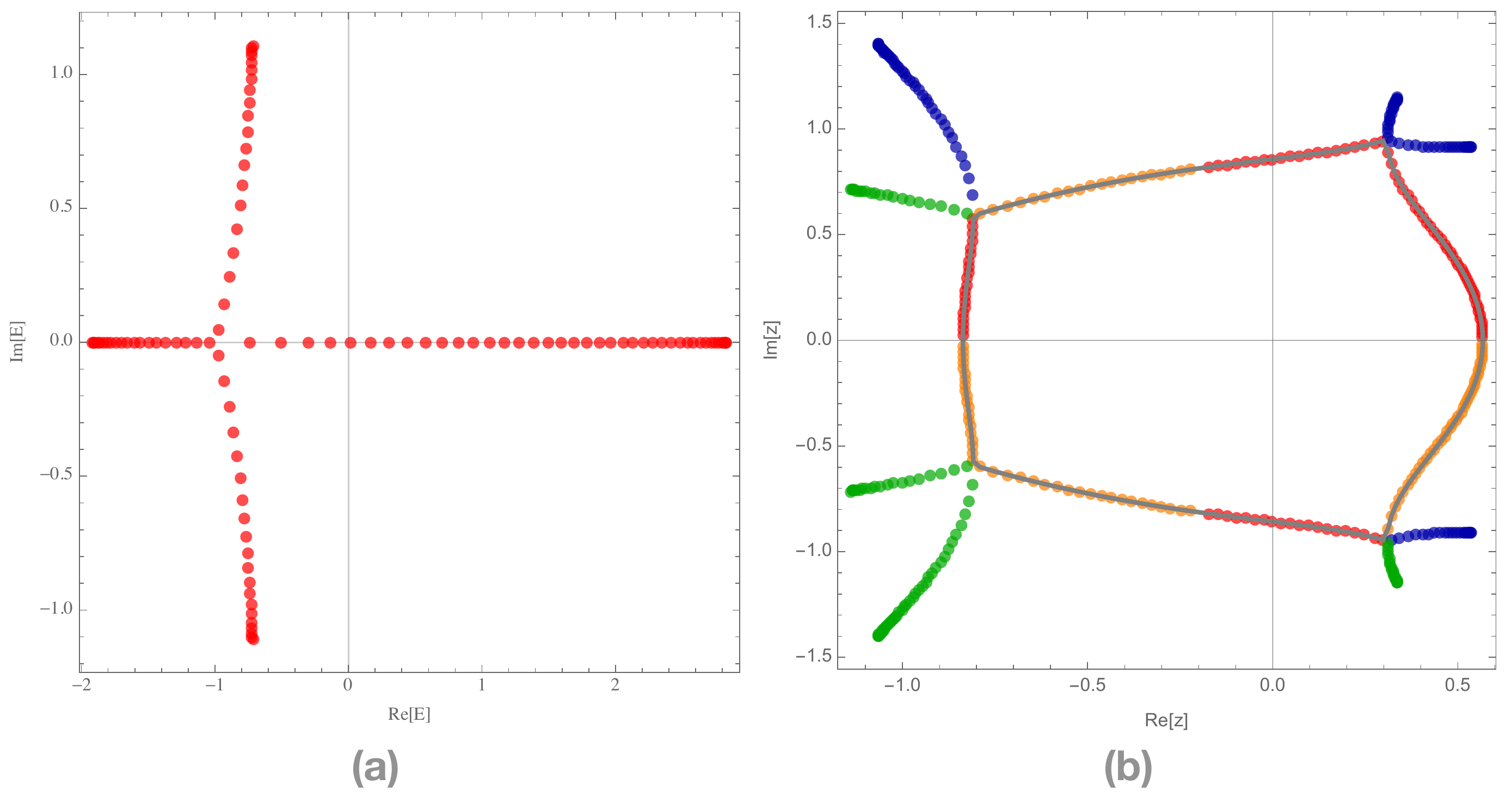}
	\par\end{centering}
	\protect\caption{\label{figs2}(a) presents eigenvalues of Hamiltonian Eq.~(\ref{a_model}) with the number of sites $L=100$ and all parameters $t_{i=-1,1,2,3}$ equal to 1. Each eigenvalue corresponds to four solutions of Eq.~(\ref{Hamz}), which are ordered by absolute value and marked in red, orange, darker blue and darker green colors respectively. 
	Then (b) shows that the first two solutions form the GBZ (gray continuous loop). }
	\end{figure}

Consider a single-band model with the following real space  Hamiltonian
\begin{equation}\label{a_model}
\hat{H}= \sum_{i=1}^L t_{-1} \hat{c}_{i+1}^{\dagger} \hat{c}_{i} + t_1 \hat{c}_{i}^{\dagger} \hat{c}_{i+1} + t_2 \hat{c}_{i}^{\dagger} \hat{c}_{i+2} + t_3 \hat{c}_{i}^{\dagger} \hat{c}_{i+3}, 
\end{equation}
and the corresponding eigenequation is
\begin{equation}\label{a_matrix}
H\Psi=E\Psi,\qquad H=\left(\begin{array}{llllllll}
{0} & {t_1} & {t_2} & t_3 & 0 & {\cdots} & {0} & {0} \\ 
{t_{-1}} & {0} & {t_1} & t_2 & t_3 & {\cdots} & {0} & {0} \\ 
{0} & {t_{-1}} & {0} & t_1 & t_2 & {\ldots} & {0} & {0} \\
0 & {0} & {t_{-1}} & {0} & t_1 & {\ldots} & {0} & {0} \\
0 & 0 & {0} & {t_{-1}} & {0} & {\ldots} & {0} & {0} \\
{\vdots} & {\vdots} & {\vdots} & {\vdots} & {\vdots} & {\ddots} & {\vdots} & {\vdots} \\ 
0 & 0 & 0 & {0} & {0} & {\ldots} & {0} & {t_1} \\
{0} & {0} & {0} & 0 & 0 & {\ldots} & {t_{-1}} & {0}\end{array}\right), \qquad \Psi=\left(\begin{array}{c}{\psi_{1}} \\ {\psi_{2}} \\ {\psi_{3}} \\ {\psi_{4}} \\ {\psi_{5}} \\ {\vdots} \\ {\psi_{L-1}} \\ {\psi_{L}}\end{array}\right).
\end{equation}
Before proceed to solve Eq.~(\ref{a_matrix}), we first review the procedure of exact solution~\cite{bottcher2005_SM,Cobanera_SM}. Recall the one-dimensional infinite square well problem in quantum mechanics. Although the translational symmetry is broken at the boundary of the well, we also need to solve the (translational invariance) Schr\"odinger equation  $(\hat{p}^2/2m+V_0)|\phi\rangle=E|\phi\rangle$. For any given $E$, there exist two linear independent plane wave solutions $|k\rangle$ and $|-k\rangle$. If their linear superposition $|\phi\rangle=c_1|k\rangle+c_2|-k\rangle$ satisfy the boundary condition, we say $|\phi\rangle$ is the eigenstate of the Hamiltonian with the  corresponding eigenvalue $E$. Back to Eq.~(\ref{a_matrix}), we first notice that the eigenequation can be separated to the {\em bulk equation} 
\begin{equation}
t_{-1}\psi_s - E\psi_{s+1}+t_1 \psi_{s+2}+t_2\psi_{s+3}+t_3\psi_{s+4}=0, 
\label{Bequation}, \quad s=1,2,...,L-4,
\end{equation}
and {\em boundary equation} 
\begin{equation}\label{split}
\begin{split}
-E \psi_1 + t_1 \psi_2 + t_2 \psi_3 + t_3 \psi_4 &= 0, \\
t_{-1}\psi_{L-3}-E \psi_{L-2}+t_1 \psi_{L-1}+t_2 \psi_L &= 0, \\
t_{-1}\psi_{L-2}-E\psi_{L-1}+t_1 \psi_{L}&=0, \\
t_1 \psi_{L-1}-E \psi_{L}&=0. 
\end{split}
\end{equation}
Here the {\em bulk equation} corresponds to the (translational invariance) Schr\"odinger equation, and the {\em boundary  equation} refers to boundary condition in the one-dimensional infinite square well problem. Since the {\em bulk equation} has discrete translational symmetry, for a given $E$, it has four linear independent eigenfunctions, which can be written as 
\begin{equation}
\Psi_i(E)=(z_i,z_i^2,...,z_i^{L-1},z_i^{L}),\quad i=1,2,3,4,
\end{equation}
where $z_i$ satisfy the following characteristic polynomial equation for given $E$
\begin{equation}\label{CEquation}
f(z_i,E) :=H(z_i)-E= t_{-1}/z_i + t_1 z_i + t_2 z_i^2 + t_3 z_i^3 - E =0, 
\end{equation}
The solution of Eq.~(\ref{a_matrix}) can be written as the linear superposition of $\Psi_i(E)$ satisfying the {\em boundary equation}. To be more precise, 
\begin{equation}\begin{aligned}
\Psi(E)&=c_1 \Psi_1(E)+c_2 \Psi_2(E)+c_3 \Psi_3(E)+c_4 \Psi_4(E)\\
&=(\psi_1,\psi_2,...,\psi_{L-1},\psi_L)^t,
\end{aligned}\end{equation}
where 
\begin{equation}\label{Gsolution}
\psi_n=\sum_{i=1}^4 c_i z_i^n 
=c_1z_1^n+c_2z_2^n+c_3z_3^n+c_4z_4^n,\quad n=1,...,L,
\end{equation}
and the solution of Eq.~(\ref{CEquation}) are ordered as follows
\begin{equation}
|z_1| \leq |z_2| \leq |z_3|\leq |z_4|.
\label{Order}
\end{equation}
Substituting Eq.~(\ref{Gsolution}) to the {\em boundary  equation} Eq.~(\ref{split}), one can obtain the following matrix equation after an appropriate transformation
\begin{equation}
H_B \begin{pmatrix}
c_1 \\ c_2 \\ c_3 \\ c_4
\end{pmatrix}=
\begin{pmatrix}
A(z_1) & A(z_2) & A(z_3) & A(z_4) \\
B_1(z_1)z_1^L&B_1(z_2)z_2^L &B_1(z_3)z_3^L&B_1(z_4)z_4^L\\
B_2(z_1)z_1^L&B_2(z_2)z_2^L &B_2(z_3)z_3^L&B_2(z_4)z_4^L\\
B_3(z_1)z_1^L&B_3(z_2)z_2^L &B_3(z_3)z_3^L&B_3(z_4)z_4^L\\ 
\end{pmatrix}
\begin{pmatrix}
c_1 \\ c_2 \\ c_3 \\ c_4
\end{pmatrix}=0, 
\label{Mform}
\end{equation}
where 
\begin{equation}\begin{aligned}
A(z_{i})&=-E z_i+t_1 z_i^2+t_2 z_i^3+t_3 z_i^4,\\
B_1(z_j)&=t_2+t_1/z_j-E/z_j^2+t_{-1}/z_j^3,\\
B_2(z_j)&=t_1-E/z_j+t_{-1}/z_j^2,\\
B_3(z_j)&=t_{-1}/z_j - E. 
\end{aligned}\end{equation}
The non-trivial solution $(c_1,c_2,c_3,c_4)$ requires 
\begin{equation}
\det[H_B] = 0. 
\label{Det}
\end{equation}
It is clear that the determinant of $H_B$ contains $4!=24$ terms, and each term is a product of elements from different rows and columns of the matrix. Hence Eq.~(\ref{Det}) can be further expressed as
\begin{equation}\begin{aligned}
&\sum_{i\neq j \neq k \neq l=1}^4 A(z_i)B_1(z_j)B_2(z_k)B_3(z_l)\times(z_j z_k z_l)^L \\
&=F_1(z,E)\times(z_2z_3z_4)^L+F_2(z,E)\times(z_1z_3z_4)^L+F_3(z,E)\times(z_1z_2z_4)^L+F_4(z,E)\times(z_1z_2z_3)^L\\
&=0,\\
\label{Detform}
\end{aligned}
\end{equation}
where the coefficient $F_i(z,E)$ is a function of $z=(z_1,z_2,z_3,z_4)$ and $E$. Since both the degrees of $A(z_i)$ and $B_{1/2/3}(z_i)$ are finite and independent of $L$, the leading term of Eq.~(\ref{Detform}) can be ordered by 
\begin{equation}
(z_2z_3z_4)^L\geq(z_1z_3z_4)^L\geq (z_1z_2z_4)^L\geq(z_1z_2z_3)^L
\end{equation}
according to Eq.~(\ref{Order}) in the thermodynamic limit. Hence in the $L\rightarrow \infty$ limit, if $|z_1(E)|<|z_2(E)|$, the only leading term of Eq.~(\ref{Detform}) is $F_1(z,E)\times(z_2z_3z_4)^L$, which requires $F_1(z,E)=0$. Since $F_1(z,E)$ is a function of $A(z,E)$ and $B_{1/2/3}(z,E)$, the order of $E$ in $F_1(z,E)$ is independent of $L$. This means there only exist finite solutions of $E$ in $F_1(z,E)=0$. As a result, they can not form a continuous spectrum as mentioned in \cite{Yokomizo2019_SM}. On the other hand, if 
\begin{equation}
|z_1(E)|=|z_2(E)|,
\label{GBZcondition}
\end{equation}
there exist two leading terms in the $L\rightarrow \infty$ limit, which implies 
\begin{equation}
\det[H_B]=0\rightarrow F_1(z,E)\times(z_2z_3z_4)^L+F_2(z,E)\times(z_1z_3z_4)^L=0.
\end{equation}
In this case 
\begin{equation}
\frac{F_1(z,E)}{F_2(z,E)}=-\left(\frac{z_1}{z_2}\right)^L. 
\end{equation}
This means the order of $E$ depends on $L$, which will form a continuous band in the thermodynamic limit. The set of $z$ satisfying Eq.~(\ref{GBZcondition}) is called generalized Brillouin zone (GBZ). From the above derivation, the GBZ condition $|z_1(E)|=|z_2(E)|$ is related to the bulk Hamiltonian 
\begin{equation}
H(z)=t_{-1}/z + t_1 z + t_2 z^2 + t_3 z^3. 
\label{Hamz}
\end{equation}
To be more precise, the order of the pole in $H(z)$ determines the form of boundary matrix $H_B$ in Eq.~(\ref{Mform}), and finally determines the condition for the continuous band Eq.~(\ref{GBZcondition}). 
In the thermodynamic limit, the spectrum as the image of GBZ on $H(z)$ is labeled by $\mathcal{L}_{GBZ}$, and needs to satisfy
\begin{equation}\label{SM_GBZcondition}
|H_m^{-1}(\mathcal{L}_{GBZ})|=|H_{m+1}^{-1}(\mathcal{L}_{GBZ})|,
\end{equation}
where $H^{-1}(E)$ is the inverse function of $H(z)$. Here we give numerical calculations in Fig.~\ref{figs2} to support the above conclusions. Next we generalize the above procedure to a general single band Hamiltonian. 

\subsection{General case} 

Consider the following general single band real space Hamiltonian, 
\begin{equation}
\hat{H}=\sum_{i,j=1}^L t_{i-j}\hat{c}_j^{\dagger}\hat{c}_{i}, 
\end{equation}
where the hopping parameters only exist in a finite range $-m \leq i - j \leq n$ and $L$ is the number of sites. For each site, the largest hopping range to left is $n$ and to right is $m$. The system reduces to Hermitian when $t_{i-j}$ equals to $t_{j-i}^*$. Similar to the derivation in the above section, this Hamiltonian can be divided into two parts: the bulk and the boundary. The bulk, ranging from $(n+1)_{th}$ site to $(L-m)_{th}$ site, maintains translational symmetry, while the boundary, including the remaining parts of two ends, has no longer the translation invariance. Starting from the eigenequation 
\begin{equation}
H \Psi  = E \Psi
\label{Eequa}
\end{equation}
we can solve the eigenequation from two parts, the bulk and boundary equations. 

\par \emph{Bulk equation:} The bulk equation is 
\begin{equation}
t_{-m} \psi_s + t_{-m+1} \psi_{s+1} + \dots + (t_0 - E)\psi_{s+m} + t_1 \psi_{s+m+1}+\dots + t_n \psi_{s+m+n} = 0, \quad s=1,...,L-(m+n),
\label{GBequation}
\end{equation}
where $\psi_s$ denotes the $s_{th}$ component of wavefunction. For the sake of simplicity, the value of $t_0$ is usually taken as zero since it does not affect the eigenfunction. For a given $E$, the bulk equation has $m+n$ linear independent eigenfunctions, which can be written as 
\begin{equation}
\Psi_i(E)=(z_i,z_i^2,...,z_i^{L-1},z_i^{L}),\quad i=1,...,m+n,
\end{equation}
where $z_i$ satisfy the following characteristic polynomial equation for given $E$
\begin{equation}
f(z_i,E) :=H(z_i)-E= \sum_{j=-m}^{n} t_j z_i^{j} - E =0.
\label{CEquation1}
\end{equation}
According to the linear superposition principle, the following wavefunction is also the eigenfunction of Eq.~(\ref{Eequa})
\begin{equation}\begin{aligned}
\Psi(E)&=\sum_{i=1}^{m+n}c_i\Psi_i(E)\\
&=(\psi_1,\psi_2, \dots ,\psi_s, \dots , \psi_{L-1},\psi_L)^t,
\end{aligned}\end{equation}
where 
\begin{equation}
\psi_s=\sum_{i=1}^{m+n} c_i z_i^s, \quad s=1,...,L,
\label{GSolution}
\end{equation}
and the solutions of Eq.~(\ref{CEquation1}) are ordered as follows
\begin{equation}
|z_1| \leq |z_2| \leq... \leq |z_{m+n-1}|\leq |z_{m+n}|.
\label{Order}
\end{equation}

\par \emph{Boundary equation:} Consider the $m+n$ boundary equations and substitute the solution Eq.~(\ref{GSolution}) into it, then one can obtain $m$ constraint equations about $\{\psi_1,\psi_2, \dots, \psi_m\}$ and $n$ limited equations about $\{\psi_{L-n+1},\psi_{L-n+2}, \dots, \psi_L\}$, where $L$ is the total number of the lattice sites. Open boundary means that there are no components of wave function beyond the two ends of the chain, namely, $\psi_{i<1}=0$ and $\psi_{i>L}=0$. In order to express the boundary equation, we first define
\begin{equation}
T_s(\psi):=t_{-m} \psi_s + t_{-m+1} \psi_{s+1} + \dots + (t_0 - E)\psi_{s+m} + t_1 \psi_{s+m+1}+\dots + t_n \psi_{s+m+n}.  
\end{equation}
Then the $m+n$ boundary equations can be expressed as 
\begin{equation}
\begin{split}
&T_s(\psi)=0,  \qquad s=-m+1, -m+2, \dots, 0; \\
&T_s(\psi)=0,  \qquad s=L-m-n+1, L-m-n+2, \dots, L-m, 
\end{split}
\end{equation}
with the open boundary condition 
\begin{equation}
\psi_{i<1}=\psi_{i>L}=0
\label{OBcondition}
\end{equation}
In fact, the form of boundary equations ensures two things. There is always at least one term, of $m+n$ terms in Eq.~(\ref{GBequation}), removed according to open boundary Eq.~(\ref{OBcondition}). And we need to make sure that the term with coefficient $E$ does not disappear. For example, $T_{-m+1}(\psi)= - E\psi_{1} + t_1 \psi_{2}+\dots + t_n \psi_{n+1} = 0 $ due to $\psi_{i<1}=0$ and $T_{L-n}(\psi)=t_{-m}\psi_{L-m} +t_{-m+1}\psi_{L-m+1} + \dots  -E \psi_L =0$ due to $\psi_{i>L}=0$. According to Eq.~(\ref{GSolution}), these $m+n$ boundary equations can be written as the following form:
\begin{equation}
H_b \begin{pmatrix}
c_1 \\ c_2 \\ \vdots \\ c_m \\ c_{m+1} \\ \vdots \\ c_{m+n}
\end{pmatrix}=
\begin{pmatrix}
A_1(z_1)  & A_1(z_2) & \dots & A_1(z_{m+n}) \\
A_2(z_1)  & A_2(z_2) & \dots & A_2(z_{m+n}) \\
\vdots & \vdots & \vdots & \vdots \\
A_m(z_1)  & A_m(z_2) & \dots & A_m(z_{m+n}) \\ 
B_1(z_1)z_1^L  & B_1(z_2)z_2^L & \dots & B_1(z_{m+n})z_{m+n}^L \\
\vdots & \vdots & \vdots & \vdots \\
B_n(z_1)z_1^L  & B_n(z_2)z_2^L & \dots & B_n(z_{m+n})z_{m+n}^L
\end{pmatrix}
\begin{pmatrix}
c_1 \\ c_2 \\ \vdots \\ c_m \\ c_{m+1} \\ \vdots \\ c_{m+n}
\end{pmatrix}=0, 
\label{GMfrom}
\end{equation}
where $z_i$ represents $i_{th}$ roots of Eq.~(\ref{CEquation1}), $A_i(z_j)$ and $B_i(z_j)$ are polynomials about $z_j$ with finite order. The matrix elements $A_i(z_j)$ can be obtained by substituting all the terms $\psi_r$ of $T_{-m+i}(\psi)$ into $z_j^r$, likely, $B_i(z_j)$ is obtained by substituting all the terms $\psi_r$ of $T_{L-m-n+i}(\psi)$ into $z_j^{r-L}$. For example, $A_1(z_j) = -E z_j + t_1z_j^2 + \dots + t_n z_j^{n+1}$ and $B_n(z_j) = t_{-m}z_j^{-m} + t_{-m+1} z_j^{-m+1} + \dots - E$.  Notice that $m,n \ll L$ and $L$ representing the size of lattice tends to infinity in the thermodynamic limit. Nontrivial solutions of Eq.~(\ref{GMfrom}) further require that 
\begin{equation}
\det[H_b]= 0. 
\label{GDet}
\end{equation}
The determinant of $H_b$, a $(m+n)*(m+n)$ square matrix, is the summation of $S_{m+n}=(m+n)!$ terms, and each term is the product of matrix elements belonging to different rows and columns. Therefore, each term contains the product of $n$ different roots $z_j^L$, and Eq.~(\ref{GDet}) can be expressed as:
\begin{equation}\begin{aligned}
&\det[H_b]\\
&=F_{1}(z,E)\times(z_{m+1}z_{m+2}...z_{m+n-1}z_{m+n})^L+F_{2}(z,E)\times(z_{m}z_{m+2}...z_{m+n-1}z_{m+n})^L+...\\
&=0,\\
\label{Detform1}
\end{aligned}
\end{equation}
where $F_{1}$ and $F_{2}(z,E)$ are the coefficients of $(z_{m+1}z_{m+2}...z_{m+n-1}z_{m+n})^L$ and $(z_{m}z_{m+2}...z_{m+n-1}z_{m+n})^L$ respectively, and the subscript $i$ of $F_i(z,E)$ increases as the magnitude of $(z_i\dots z_j)^L$ gradually decreases. In this way,  $F_1(z,E)$ always corresponds to the term with the largest magnitude, namely, $(z_{m+1}z_{m+2}...z_{m+n-1}z_{m+n})^L$, and so on. Based on the similar reasons with the previous part, the continuous band requires 
\begin{equation}
|z_m(E)|=|z_{m+1}(E)|, 
\end{equation}
such that the leading order of Eq.~(\ref{Detform1}) is equivalent to 
\begin{equation}
\left(\frac{z_{m+1}}{z_m}\right)^L=-\frac{F_{1}(z,E)}{F_{2}(z,E)}. 
\label{GConstraint}
\end{equation}
In fact, Eq.~(\ref{GConstraint}) gives the constraint condition on $E$ because $z$ is the function of $E$. The exact solutions of $|z_m(E)|=|z_{m+1}(E)|$ give the spectrum in the thermodynamic limit and corresponding $z$ forms the generalized Brillouin zone. The right side of Eq.~(\ref{GConstraint}) includes the detailed information about open boundary while the left side does not. It means that boundary conditions have little effect on the properties of the bulk if $N$ tends to infinity, but we need pay special attention to the fact that the above statements are based on the open boundary, it is no longer true for nonlocal boundary terms, such as periodic boundary. 


\subsection{Two-band model with sublattice symmetry}

The above discussions can be naturally extended to multi-band cases, especially two-band Hamiltonian with sublattice symmetry. Generically, the bulk Hamiltonian is written as 
\begin{equation}
H(z) = 
\begin{pmatrix}
0 & \frac{P^{(1)}_{m_1+n_1}(z)}{z^{m_1}} \\
\frac{P^{(2)}_{m_2+n_2}(z)}{z^{m_2}} & 0 
\end{pmatrix}. 
\label{ChiralH}
\end{equation}
The off-diagonal terms of the matrix are two different functions that are holomorphic at the entire complex plane except the origin. Next we show that the commonly used two-band tight-binding model can always map to the single-band model due to the presence of sublattice symmetry. According to Eq.~(\ref{ChiralH}), one can immediately write down the characteristic equation, 
\begin{equation}\label{a_twoce}
F(z,E) = 
\frac{P_{m+n}(z)-E^2 z^{m}}{z^{m}} = 0, 
\end{equation}
where $m=m_1+m_2$ and $n=n_1+n_2$, and this equation establishes the mapping between $z$ and $E$. Additionally, sublattice symmetry ensures that the eigenvalues $(E,-E)$ always appear in pairs. Therefore, it comes to the conclusion that one can always take the base energy as $E^{2}$ in single-band model, which corresponds to the $\pm E$ in two-band Hamiltonian with sublattice symmetry, and they have the same characteristic equation that determines the generalized Brillouin zone. Then we can get the same conclusion as the single-band model 
\begin{equation}\label{a_twogbz}
|z_{m}(E)|=|z_{m+1}(E)|, 
\end{equation}
where $m$ represents the order of the pole of characteristic equation Eq.~(\ref{a_twoce}), and this conclusion is confirmed numerically in Fig.~\ref{figs5}. 

\section{GBZ encircles the same number of zeros and poles}\label{apdnx:3}
In this section, we formally prove that in single-band cases, GBZ encloses the same number of zeros and poles of the characteristic polynomial. Here each zero and pole are counted as many times as its multiplicity and order, respectively. Specifically, if GBZ is a simple closed curve (Jordan curve), the same conclusion can also be proved by using the Rouché$^{\prime}$s theorem. 

\subsection{Single-band GBZ}
For the single-band characteristic equation $E=H(z)$, given a value of $E$, the corresponding $m+n$ zeros can be obtained, where $m$ is the order of the pole. In this section, we prove that the single-band GBZ is just the boundary of the open set that consists of the first $m$ zeros for all $E\in \mathbb{C}\setminus \mathcal{L}_{GBZ}$. As a result, the GBZ is a closed curve that encircles the same number of zeros and poles. Reasonably, we conjecture that this closed curve (or single-band GBZ) should be connected. 

\subsubsection{Definition and characteristics of the open set of zeros}

In order to come to the final conclusions about single-band GBZ, we first give the following preliminary knowledge. The (single-band) characteristic equation $f(z,E)=E-H(z)=0$ establishes a mapping from $z$ to $E$. For a given $z$, only a unique $E$ can be obtained. Conversely, given a value of $E$, we will obtain corresponding $(m+n)$ zeros sprinkled on the complex plane, which can be marked by different colors according to their magnitude. As we sweep $E$ through the entire complex plane, we can obtain a set of continuum areas, labeled by $\mathcal{A}_{1},\mathcal{A}_{2},\dots,\mathcal{A}_{m+n}$, whose colors are different, as shown in Fig.~\ref{figs4}(d). 

Here we take a strict definition for the open set $\mathcal{A}_i$. Firstly, order the roots of the characteristic polynomial by their magnitude as $|z_1(E)|\leq|z_2(E)|\leq\cdots\leq|z_{m+n}(E)|$, then the open set $\mathcal{A}_i$ is defined as
\begin{equation}
\mathcal{A}_i:=\{ z_i\in \mathbb{C}|\forall E\in \mathbb{C}:|z_{1}(E)|\leq\dots\leq |z_{i-1}(E)|<|z_i(E)|<|z_{i+1}(E)|\leq\dots\leq|z_{m+n}(E)| \},
\label{Ai}
\end{equation}
from which we can see that $\mathcal{A}_i$ is the subset of $\mathbb{C}$, i.e., $\mathcal{A}_i\subset \mathbb{C}$. 
Now suppose that there is an intersection between $\mathcal{A}_i$ and $\mathcal{A}_j$, which is labeled by $I_{ij}$.  Then $I_{ij}$ is also an open set because any finite intersection of open sets is an open set. Note that the element of $I_{ij}$ must be the element in $\mathcal{A}_i$ and $\mathcal{A}_j$ simultaneously. Hence, for any point $z_0\in I_{ij}$, we have $z_0=z_i(E)=z_j(E')$ for $E\neq E'$. We note that according to the definition of Eq.~\ref{Ai}, $E$ can not be equal to $E'$. On the other hand, according to $z_0=z_i(E)=z_j(E')$ for $E\neq E'$, 
the same $z_0$ is mapped to two different values $E\neq E^{\prime}$. That is contrary to the injective function $H:z\rightarrow E$ in the single-band systems, hence 
the open sets $\mathcal{A}_i$ and $\mathcal{A}_j$ have no intersection in single-band systems, or geometrically speaking, $\mathcal{A}_i$ and $\mathcal{A}_j$ have no overlap. We finally note that this property only works for single-band models. 



\subsubsection{GBZ is the boundary of the open set consisting of the first $m$ zeros}

\begin{figure}[t]
	\begin{centering}
	\includegraphics[width=0.8\linewidth]{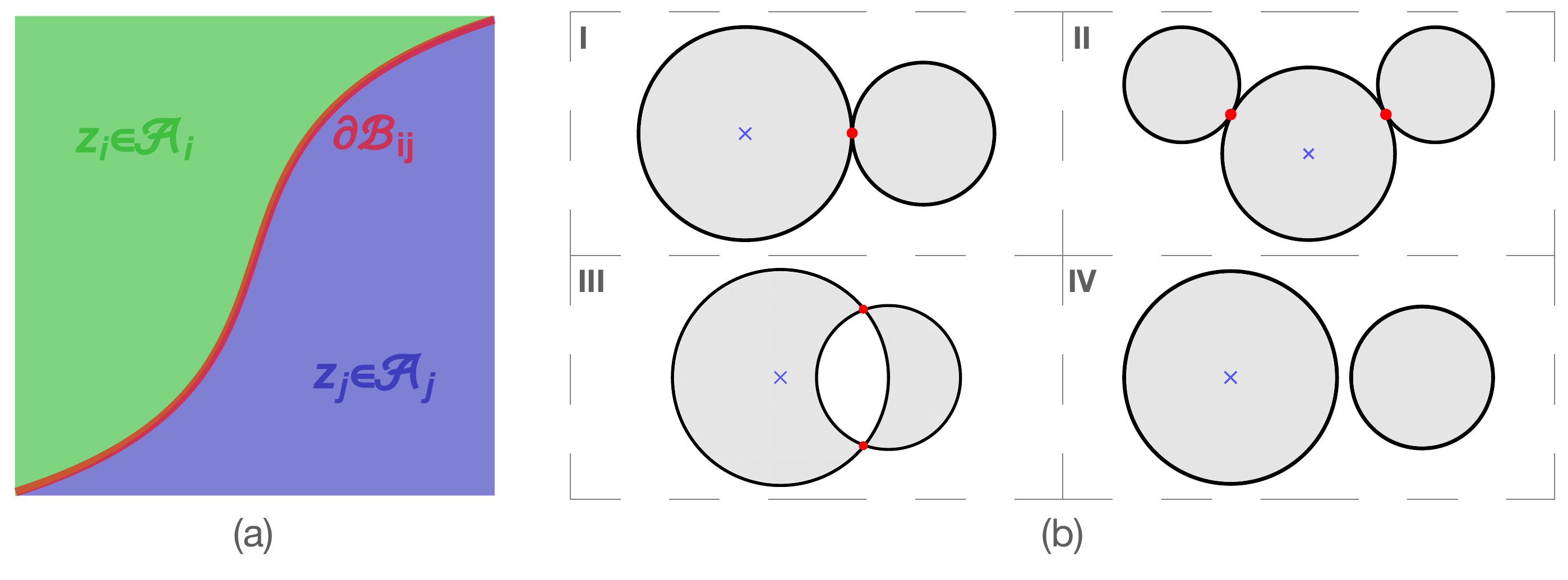}
	\par\end{centering}
	\protect\caption{\label{figs6}(a) plots the distribution of the zeros of the characteristic polynomial, in which the $i$-th zero after ordering the zeros belongs to the area $\mathcal{A}_i$, and $\partial\mathcal{B}_{ij}$ represents the boundary between the two areas $\mathcal{A}_i$ and $\mathcal{A}_j$. (b) gives some possible geometrical configurations of GBZ on the complex plane. The red dots represent the self-intersections, and the blue crosses denote the pole (at the origin of the complex plane). The solid black curve represents GBZ, and the gray shadow area represents the interior region of GBZ.}
\end{figure}

We now define the boundary between the open sets $\mathcal{A}_i$ and $\mathcal{A}_j$. Due to $\mathcal{A}_i$ being the subset of $\mathbb{C}$, the complementary set is $\mathcal{A}_i^c:=\mathbb{C}\setminus \mathcal{A}_i$. The frontier of $\mathcal{A}_i$ is the set of points $z\in\mathbb{C}$ such that every neighborhood of $z$, which is labeled by $O$, contains at least one point of $\mathcal{A}_i$ and at least one point not of $\mathcal{A}_i$. Mathematically, we can  express the frontier of $\mathcal{A}_i$ as follows 
\begin{equation}
	\partial \mathcal{A}_i := \{z\in\mathbb{C}|\forall O\ni z:O\cap \mathcal{A}_i\neq 0 \ {\rm and}\ O\cap \mathcal{A}_i^c\neq 0\}.
\end{equation}
Then the boundary between $\mathcal{A}_i$ and $\mathcal{A}_j$ is defined as 
\begin{equation}
	\partial \mathcal{B}_{ij}=\partial \mathcal{A}_i \cap \partial \mathcal{A}_j. 
\end{equation}
The points on the boundary $\partial \mathcal{B}_{ij}$ is called boundary points, which do not belong to the open set $\mathcal{A}_i$ and $\mathcal{A}_j$. Hence the set $\partial \mathcal{B}_{ij}$ can be expressed as by definition Eq.(\ref{Ai})
\begin{equation}
\partial \mathcal{B}_{ij}=\{z_i,z_j\in\mathbb{C}|\forall E\in\mathbb{C}:|z_i(E)|=|z_j(E)| \}. 
\end{equation}
On the other hand, due to the injectivity of $H$, $\mathcal{A}_i$ and $\mathcal{A}_j$ have no overlap for any $i$ and $j$. This results the geometrical picture illustrated in Fig.~\ref{figs6}(a). 

Review that the single-band GBZ is a set expressed as 
\begin{equation}
	GBZ:=\{z_m,z_{m+1}\in\mathbb{C}|\forall E\in\mathbb{C}:|z_m(E)|=|z_{m+1}(E)|\},
\end{equation}
where $m$ denotes the order of the pole of the characteristic equation. Obviously, the intersection between GBZ and the open sets $\mathcal{A}_{m}$ and $\mathcal{A}_{m+1}$ is an empty set by definition. Now we define the set $\mathcal{U}$ as 
\begin{equation}
	\mathcal{U}:=\{ z_1,z_2,\dots,z_m\in \mathbb{C}|\forall E\in \mathbb{C}:|z_{1}(E)|\leq\dots\leq|z_m(E)|<|z_{m+1}(E)|\leq\dots\leq|z_{m+n}(E)| \},
\end{equation}
which means that the set $\mathcal{U}$ is the union set of open sets $\mathcal{A}_1,\mathcal{A}_2,\dots, \mathcal{A}_m$ and their boundary $\partial \mathcal{B}_{ij}$ ($1\leq i,j \leq m$). Obviously, $\mathcal{A}_m$ must be the subset of $\mathcal{U}$, and the boundary $\partial\mathcal{B}_{m,m+1}$ between $\mathcal{A}_m$ and $\mathcal{A}_{m+1}$ is just the boundary $\partial \mathcal{U}$ between $\mathcal{U}$ and $\mathcal{A}_{m+1}$. Next we prove that the boundary $\partial\mathcal{B}_{m,m+1}$ is just the GBZ. 

If $z\in \partial\mathcal{B}_{m,m+1}$, then it must satisfy $|z_m(E)|=|z_{m+1}(E)|$,  which is just the definition of GBZ.  Here we notice that the boundary $\partial\mathcal{B}_{m,m+1}$ is also the boundary between $\mathcal{U}$ and $\mathcal{A}_{m+1}$. In summary, we come to the conclusion that the boundary $\partial \mathcal{U}$ of the open set $\mathcal{U}$ is exactly the GBZ, which must be a closed curve. 


We note that according to the definition of $\mathcal{U}$, its interior must automatically contains the first $m$ zeros for any $E\in\mathbb{C}$. As a result, the boundary $\partial \mathcal{U}$, which is GBZ, must always encloses the $m$ zeros and the pole of order $m$. This completes the proof. 

\subsubsection{The conjecture about the connectedness of single-band GBZ}

Now we argue and conjecture that the boundary $\partial \mathcal{U}$ (GBZ, a closed curve) should also be connected in single-band cases. For single-band case, the map $H:z\rightarrow E$ and its inverse map $H^{-1}$ are all continuous maps. If $E$ is mapped to the two points $z(E)$ and $z'(E)$ on the GBZ that satisfy $|z(E)|=|z'(E)|$, then any energy point in the neighborhood of $E$ is also mapped to two roots that are in the neighborhoods of $z(E)$ and $z'(E)$ respectively. Therefore, as $E$ varies adiabatically, the corresponding two roots $z(E)$ and $z'(E)$ change continuously. 

We assume that GBZ is composed of several disconnected parts as illustrated in case IV of Fig.~\ref{figs6}(b), then there are two different cases. In one case, $z(E)$ and its partner $z'(E)$ are always on a connected part of GBZ for all $E$, then several disconnected GBZ parts correspond to several disjoint continuum bands respectively, which contradicts the connectedness of the spectrum of the single-band Hamiltonian. As a result, the first case is also impossible. In the other case, $z(E)$ and its partner $z'(E)$ are respectively in the disconnected GBZ parts for all $E$. However, there always exist the branch points $E_0$ such that the characteristic polynomial $f(E_0,z)$ has multiple roots $z(E_0)=z^{\prime}(E_0)$, which means that the two roots $z(E_0)$ and $z^{\prime}(E_0)$ must be on the same position of GBZ. The contradiction appears, hence the second case is also impossible. To sum up, if GBZ consists of disconnected parts, contradiction always exists. Therefore, single-band GBZ must be a connected curve. 

\subsubsection{Discussion and conclusion}
We roughly divide the single-band GBZ into two cases: simple closed curves and other more complex set-ups. For the cases where single-band GBZ is a simple closed curve (Jordan curve), a more intuitive discussion is presented in the Sec.~III.~C.
For other complex cases, such as the presence of self-intersections on GBZ, we list some possible geometrical configurations of GBZ in Fig.~\ref{figs6}(b). Note that the gray shadow areas in Fig.~\ref{figs6}(b) represent the interior region of GBZ. The characteristic equation of single-band systems can be written as $E=H(z)$, and $H$ is an injective function that maps $z$ to $E$. We select the interior region (that is just $\mathcal{U}$) of GBZ that always includes the first $m$ zeros for any $E\in \mathbb{C}\setminus \mathcal{L}_{GBZ}$, where the zeros are ordered by the absolute values, i.e., $|z_1(E)|\leq \dots \leq|z_m(E)|\dots \leq|z_{m+n}(E)|$. Hence, any point $z$ in the shadow areas satisfies $|z|<|H^{-1}_{m+1}(H(z))|\leq\dots\leq|H^{-1}_{m+n}(H(z))|$, where $H^{-1}_{m+n}(H(z))$ are the $m+n$ branches of the inverse function of $H(z)$. Particularly, for case III in Fig.~\ref{figs6}(b), there are many choices for the orientation of GBZ. However, once we identify the interior of GBZ, its orientation is uniquely determined. 

Based on the above discussions, we naturally come to the conclusion that each part loop of single-band GBZ includes a different number of zeros, and the summation of the number of zeros in all part loops equals to the order of the pole. Finally, in single-band cases the winding number of open-boundary spectrum with respect to base energy $E_b\notin \mathcal{L}_{GBZ}$ can be obtained 
\begin{equation}
	w_{\mathrm{GBZ},E_b}=\frac{1}{2\pi}\sum_i\oint_{C_i}\frac{d}{dz}\arg[H(z)-E_b]dz=\sum_{i}N_{z,i}-N_p=0,
\end{equation}
where $N_{z,i}$ denotes the number of zeros included by the $i$-th part loop $C_i$ of single-band GBZ, and $N_p$ represents the order of the pole. 

\begin{figure}
	\begin{centering}
	\includegraphics[width=0.7\linewidth]{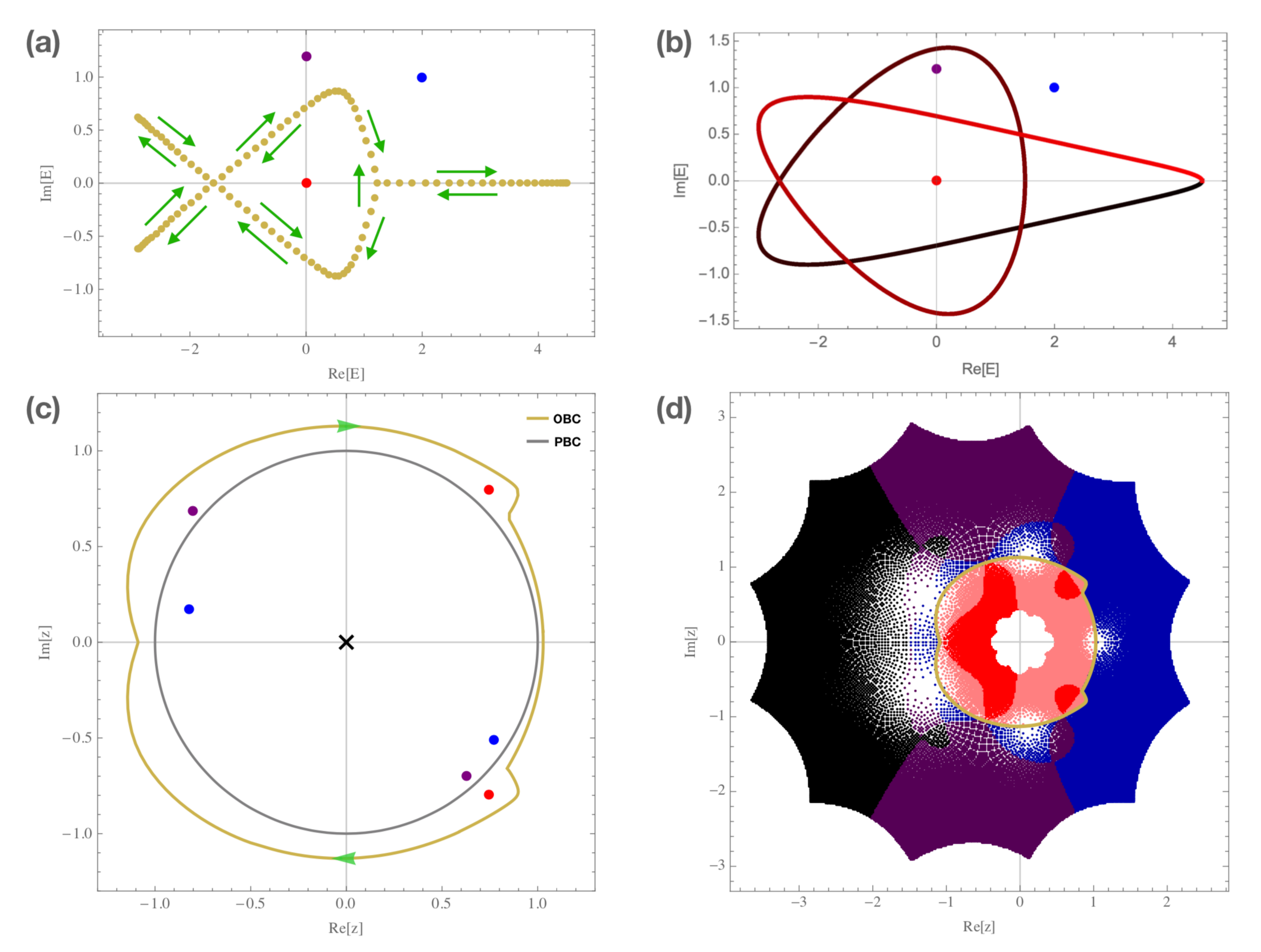}
	\par\end{centering}
	\protect\caption{\label{figs4}(a) represents the energy spectrum of the non-Hermitian system determined by Hamiltonian Eq.~(\ref{b_ham1}) under open boundary condition with parameters $ \{t_3,t_2,t_1,t_{-1},t_{-2}\}=\{1/2, 1, 1/2, 1/2, 2 \}$ and the number of site $L=110$, and the green arrows denote the direction of energy band as $z$ clockwise along GBZ depicted in (c); 
	(b) shows the energy spectrum with periodic boundary and the color gradually turns from black to red as k change from $0$ to $2\pi$. Given three base energy $E_1=0$ in red color, $E_2=1.2 I$ in purple color and $E_3=2+I$ in blue color, and their corresponding zeros are colored with the same color on the $z$ plane; 
	In Fig.~(c), the gray unit circle and yellow loop indicate GBZ of the system with periodic and open boundary respectively, and the cross notation denotes the pole with order two. Here only the first two, with smaller absolute values, of the five zeros are depicted; 
	(d) presents the distribution of five roots of Eq.~(\ref{b_ham1}) as $E$ varying uniformly from $-10-10I$ to $10+10I$. The red and pink represent the first two roots, blue, purple and black areas represent the last three roots respectively. GBZ happens to be the boundary between the second root (the pink area) and the third root (the blue area). }
	\end{figure}
\subsection{Example 1: Single-band case}
Here we take a special single-band model to further confirm numerically the main conclusions, and the bulk Hamiltonian is shown as: 
\begin{equation}\label{b_ham1}
H(z)= \sum_{i=-2}^3 t_i z^i, 
\end{equation}
where $z$ is complex variable and $t_i$ denotes the hopping parameter. Then transform Eq.(\ref{b_ham1}) into the form, $(t_3 z^5 + t_2 z^4 + t_1 z^3 - E_b z^2 + t_{-1} z + t_{-2})/z^2=0$. We fix the other coefficients, then let $E_b$ arbitrarily take values in the complex plane except for the energy spectrum $\mathcal{L}_{GBZ}$, and always get five zeros, two of which are inside GBZ and the rest are outside it, which is illustrated in Fig.~\ref{figs4}. It further comes to a conclusion that if and only if GBZ is the boundary of the region composed of the first two solutions ordered by absolute value, the winding regard to any choices of base energy $E_b$ vanish. 

In Fig.~\ref{figs4}, We have chosen three representative base energy $E_b$, which are marked in different colors, and the winding with respect to base energy $E_1, E_2, E_3$ equal to $-2,-1,0$ respectively under periodic boundary conditions, as well as all of which equal to zero with open boundary. It is worth noting that even if the spectrum surrounds the $E_1$(the red dot) in Fig.~\ref{figs4}(a), the winding of the energy is still zero. The reason is that as z moves clockwise along GBZ, its image $\mathcal{L}_{GBZ}$ keeps “back-stepping” itself and has no interior. These following significant conclusions have been confirmed numerically: 
(i:) Once we have determined the order m of the pole of $H(z)$, GBZ always contains m zeros accordingly, regardless of how the base energy takes values on the complex plane except for $\mathcal{L}_{GBZ}$.  
(ii:) If $z$ directionally circles BZ, the corresponding $\mathcal{L}_{BZ}$ will surround a finite area, while if $z$ moves along GBZ, $\mathcal{L}_{GBZ}$ always wraps around itself and contains zero area, which refers to the collapse from $\mathcal{L}_{BZ}$ to $\mathcal{L}_{GBZ}$. 
(iii:) For each $E_b$, there are always $(m + n)$ zeros distributed on the complex plane. As $E_b$ takes values throughout the complex plane except for the energy spectrum $\mathcal{L}_{GBZ}$, the boundary between the corresponding first $m$ zeros and the last $n$ zeros happens to be the GBZ curve.

\subsection{The case where GBZ is a Jordan curve}
In this subsection we focus on the cases where GBZ is a simple closed loop (Jordan curve) and the characteristic equation has the form $f(z,E)=(P_{m+n}(z)-E z^m)/z^m $. We strictly prove that GBZ encloses the same number of zeros and poles, consequently, the winding number of open-boundary spectrum is always zero. 

Assume that there is a simple closed loop $C_{z}$ in the complex plane, which can be mapped into the set $\mathcal{L}_{C_z}$ living in the complex energy plane by $f(z,E)$. If one chooses the base energy $E_b\in\mathbb{C}\setminus \mathcal{L}_{C_z}$, then one can find corresponding $(m+n)$ zeros of $f(z,E)=0$, which are located on the $(\Re(z), \Im(z))$ coordinate system. We present the following lemma before coming to the final conclusion. 

\subsubsection{Lemma}

\textbf{\textit{Lemma:}}  If there exists a smooth path $E_s$ on the complex plane, which is controlled by the parameter $s$ $(0\leq s \leq 1)$ and has no intersections with the set $\mathcal{L}_{C_z}$. Then any two points connected by this path $E_s$ have the same number of zeros of $H(z)-E_s=0$ within the closed loop $ C_z $. 

This lemma is rigorously proven in the following. 

\textbf{\textit{Proof:}} Our proof is based on the Rouché$^{\prime}$s theorem, which states that for any two complex functions $f$ and $g$ holomorphic inside the region surrounded by closed contour $\mathcal{C}_z$, if $|g(z)|<|f(z)|$ on $\mathcal{C}_z$, then $f$ and $f+g$ have the same number of zeros inside $\mathcal{C}_z$. The theorem assumes that $\mathcal{C}_z$ is a simple closed curve without self-intersections.  

The zeros of characteristic equation are determined by $P_{m+n}(z)-z^m E_b=0$. Defining the analytic functions of $z$ as
\begin{equation}
\begin{split}
	&g_s(z):=P_{m+n}(z)-z^m E_s,
\end{split}
\end{equation}
where the subscript $s\in[0,1]$. If $E_s$ changes to $E_{s+\delta}$, the equation can be expressed as 
\begin{equation}
	g_{s+\delta}(z):=P_{m+n}(z)-z^m E_{s+\delta}.
\end{equation}
Hence one can get a series of functions $g_{s}(z)$ of $z$ as $s$ varies from $0$ to $1$, and here $g_s(z)$ and $g_{s+\delta}(z)$ are just two characteristic equation with different base energy $E_s$ and $E_{s+\delta}$. Note that $E_s$ does not belong to $\mathcal{L}_{C_z}$, if it does, then some roots of $ g_s(z)$ must be on the loop $C_z$ and the Rouché$^{\prime}$s theorem is not applicable. As $\delta$ approaches zero, $E_s$ and $E_{s+\delta}$ become very close and $|E_{s+\delta}-E_s|$ tends to zero. On the other hand, for any point $z_c\in C_z$, from  $|g_s(z_c)|=|P_{m+n}(z_c)-z_c^mE_s|=|z_c^m(\mathcal{L}_{z_c}-E_s)|$, one can always find a small enough $\delta$ such that the following equation is satisfied
\begin{equation}
|g_s(z_c)| = |(\mathcal{L}_{z_c}-E_s)z_c^m|> |g_{s+\delta}(z_c)-g_s(z_c)|=|(E_{s}-E_{s+\delta}) z_c^m|. 
\end{equation}
According to Rouché$^{\prime}$s theorem, $g_s(z)$ and $g_{s+\delta}(z)$ have the same number of zeros in the area surrounded by $C_{z}$. Based on the same argument, $g_{s+\delta}(z)$ and $g_{s+\delta+\delta}(z)$ must also have the same number of zeros inside $C_z$. Thus it comes to the conclusion that any two energies $E_0$ and $E_1$, connected by a path that does not intersect with $\mathcal{L}_{C_z}$, must have the same number of zeros within $C_z$. Here the lemma has been proved. 
\begin{figure}
	\begin{centering}
	\includegraphics[width=0.7\linewidth]{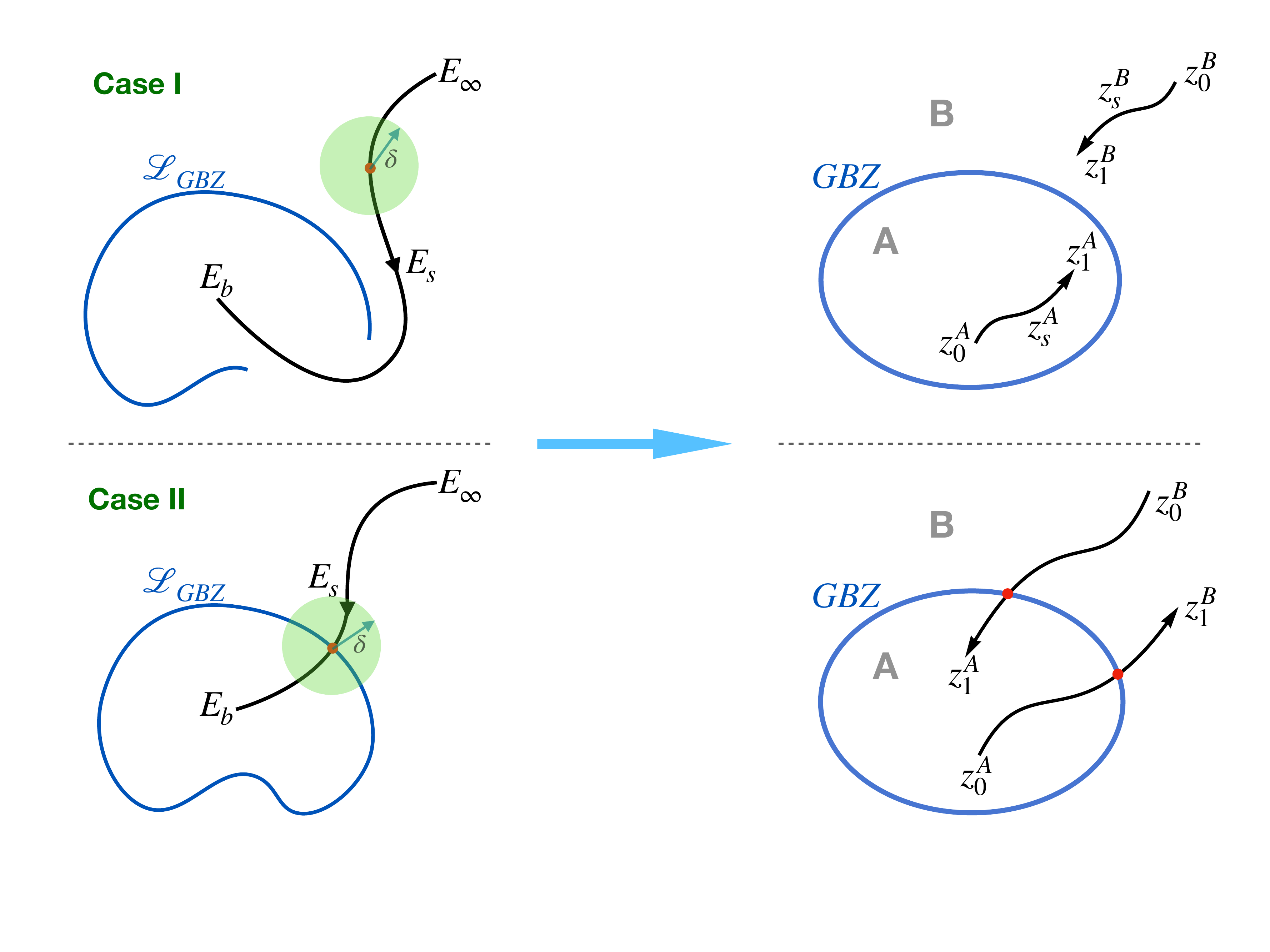}
	\par\end{centering}
	\protect\caption{\label{figs3}An illustration of the proof. The $z_s^{A/B}$ curves represent the loci of $z$ roots of $ P_{m+n}(z) - z^m E_s$. As parameter $s$ changes from $0$ to $1$, the energy changes from $ E_{\infty} $ to arbitrary chosen base energy $ E_b $, and the corresponding zeros move continuously from $ z_0^{A/B} $ to $ z_1^{A/B} $. There are two different cases. In case I, we always choose one locus of $ E_s $ that has no intersections with the open-boundary spectrum $\mathcal{L}_{GBZ}$, hence the corresponding zeros will not enter or escape from GBZ such that the total number of zeros inside GBZ is invariant as $E_s$ changes from $s=0$ to $s=1$. In case II, where $\mathcal{L}_{GBZ}$ is a geometrically closed-loop on the complex plane, the trajectory from $E_{\infty}$ to $E_b$ must intersect with $\mathcal{L}_{GBZ}$. Accordingly, due to the limitation of GBZ condition $|z_{m}(\mathcal{L}_{GBZ})|=|z_{m+1}(\mathcal{L}_{GBZ})|$, one zero inside GBZ goes out, and another zero outside GBZ comes in, such that the total number of zeros within GBZ is still unchanged.}
\end{figure}

\subsubsection{Applying the Lemma to the GBZ}

We now apply the lemma to non-hermitian systems, in which the closed loop $C_z$ represents GBZ. Notice that the GBZ is formed by the $m$-th and $(m+1)$-th zeros of characteristic equation for eigenvalues, expressed as $|z_{m}(\mathcal{L}_{GBZ})|=|z_{m+1}(\mathcal{L}_{GBZ})|$, where $m$ indicating the order of the pole. According to the geometry of the energy spectrum $\mathcal{L}_{C_z}$ on the complex plane, there are two differences cases as shown in Fig.~\ref{figs3}. 

In one case, the spectrum $\mathcal{L}_{GBZ}$ is not a closed loop geometrically, then one can always find a path $E_s$ that connects arbitrary two energies on the complex plane but has no intersections with $\mathcal{L}_{GBZ}$. Here we take $E_0$ as $E_{\infty}$ and take $E_1$ as the chosen base energy $E_b$, as shown in Fig.~\ref{figs3}, where $E_b\notin \mathcal{L}_{GBZ}$, then $E_b$ has the same number of zeros as $E_{\infty}$ within the area enclosed by GBZ. For single band system, the characteristic equation can be written as
\begin{equation}
	t_{-m}/z^{m}+t_{-m+1}/z^{m-1}+\dots + t_{-1}/z+ t_1 z+\dots + t_{n} z^{n} = E,
\end{equation}
where $t_{i=-m,\dots,n}$ represent hopping parameters. For each given energy $E$, corresponding $(m+n)$ zeros and $m$ poles at the origin can be obtained. As $E$ tends to infinity, there are $m$ zeros verge to origin and $n$ roots tend to infinity, which means that there are $m$ zeros inside the GBZ and are the same as the number of poles. Any two energies connected by $E_s$ have the same number of zeros inside GBZ, therefore, there are always the same number of zeros and poles inside GBZ for any base energy $E_b\notin \mathcal{L}_{GBZ}$, just as illustrated by case I in Fig.~\ref{figs3}. In the upper right subfigure of Fig.~\ref{figs3}, once the $E_s$ is varying, all the zeros with ordering lesser than m only moves inside the GBZ, while the ones with ordering larger than m moves outside the GBZ.

In another case, the spectrum $\mathcal{L}_{GBZ}$ forms closed loop as shown in case II in Fig.~\ref{figs3}, which divides the complex plane into two disconnected areas. The energies in each connected region have the same number of zeros inside GBZ. Now we show that any two energies belonging to distinct regions can also have the same number of zeros inside the GBZ. Now consider an arbitrary path from $E_{r-\delta}(0<r<1,\delta>0)$ to $E_{r+\delta}$ though $E_{r}$ that is exactly on $\mathcal{L}_{GBZ}$. As $\delta$ from $r$ to $1-r$, $E_{r-\delta}$ moves from $E_{\infty}$ to $E_b$ as illustrated in case II of Fig.~\ref{figs3}. Since $E_{r-\delta}$ connects to infinity, there are exactly $m$ roots of $H(z)-E=0$ that lie in GBZ. If we order the zeros, then the $m$-th zero, $z_m(E_{r-\delta})$ is inside GBZ, and the $m+1$-th zero $z_{m+1}(E_{r-\delta})$ is outside. Now we move from $E_{r-\delta}$ to $E_{r+\delta}$. Due to the property of GBZ, $|z_{m}(E_r)|=|z_{m+1}(E_r)|$, both $z_m(E_r)$ and $z_{m+1}(E_r)$ are right at GBZ. So we have in our mind a picture of $z_m(E)$ moving towards GBZ from inside, while $z_{m+1}(E)$ moving towards GBZ from outside, as $E$ moves from $E_{r-\delta}$ to $E_{r}$, crossing the GBZ at two different points (the two red dots in the lower right subfigure in Fig.~\ref{figs3}) when $E=E_r$. Consequently, in the process of $E_s$ passing through the energy spectrum $\mathcal{L}_{GBZ}$, we can always find a path $E_s$ such that there is always one zero from the inside of GBZ to the outside, and another zero from the outside to inside, so as to ensure that the number of zeros enclosed by GBZ remains unchanged and are the same as the order of the pole. 

In summary, we obtain that GBZ always encircles the same number of zeros and poles for any base energy $E_b\notin \mathcal{L}_{GBZ}$. Thus the winding number of energy spectrum with open boundary is calculated
\begin{equation}
w_{GBZ,E_b}= \frac{1}{2\pi i} \oint_{C_z} \frac{H'(z)}{H(z)-E_b} d z = N_{zeros}-N_{poles} = 0  
\end{equation}
where $N_{zeros}$ denotes the number of zeros and $N_{poles}$ the number of the poles. 

\subsection{Example 2: Two-band case}
\begin{figure}
	\begin{centering}
	\includegraphics[width=0.7\linewidth]{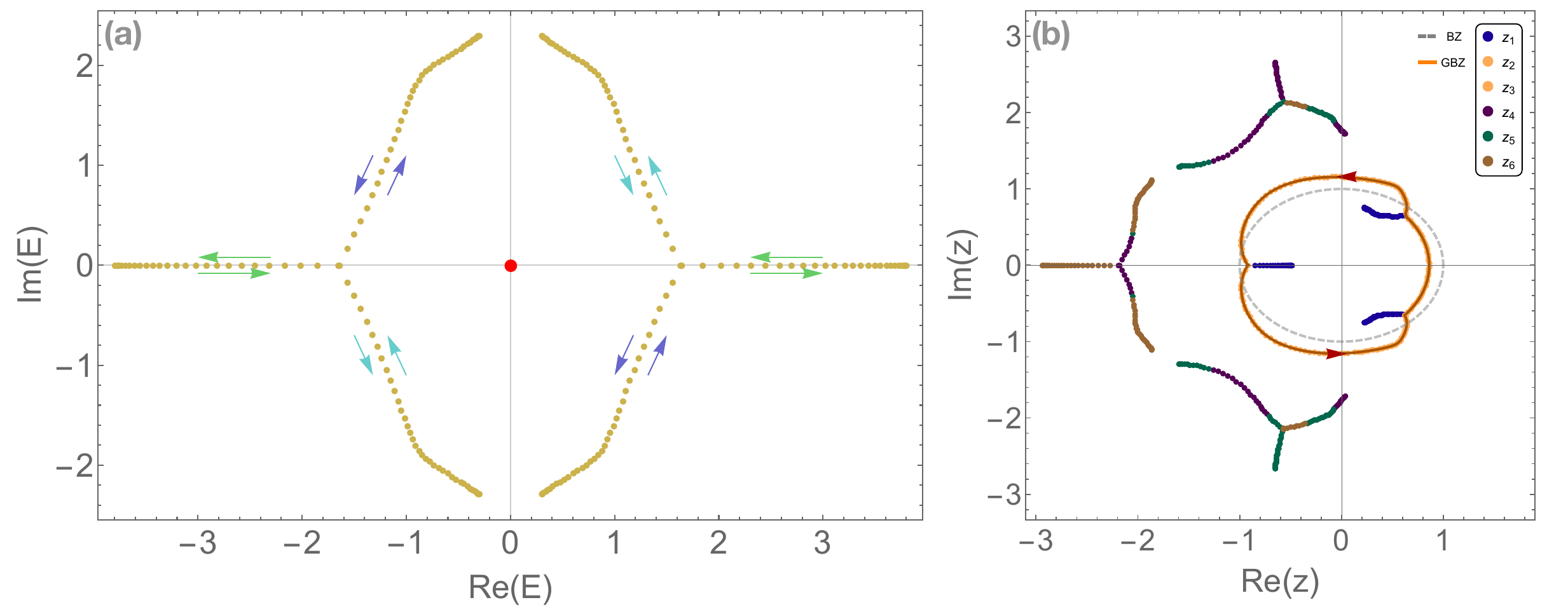}
	\par\end{centering}
	\protect\caption{\label{figs5}(a) illustrates the eigenvalues of non-Hermitian system determined by Hamiltonian Eq.(\ref{S20}) under open boundary condition with parameters $ \{t_1,t_2,t_3,w_1,w_2,w_3\}=\{1,4,1,1,1,\frac{1}{2}\}$ and size $L=75$, and the red point denotes two-degenerate edge state protected by sublattice symmetry. The arrows represent the evolution orientation of eigenvalues as $z$ anticlockwise along the GBZ; 
	(b) shows the GBZ in orange color and BZ in dashed gray color, and the roots $z_i$, obtained by substituting eigenvalues into $H^{-1}(E)$, are ordered in ascending amplitude and marked in different colors. }
\end{figure}
Consider a two-band Hamiltonian with respect to sublattice symmetry in real-space representation, which is shown as:
\begin{equation}
H = \sum_{i=1}^L t_1 a_i^{\dagger}b_i + t_2 a_{i+1}^{\dagger}b_i + t_3 a_i^{\dagger}b_{i+1} + w_1 b^{\dagger}_{i+1}a_i+w_2 b_i^{\dagger}a_{i+2} + w_3 b_i^{\dagger}a_{i+3}
\end{equation}
with the number of unit cell $L$ and two degrees per unit cell. Then the corresponding bulk Hamiltonian is written as: 
\begin{equation}
H(z)=
\begin{pmatrix}
0 & \frac{t_2+t_1 z +t_3 z^2}{z} \\
\frac{w_1+w_2 z^3+w_3 z^4}{z} & 0
\end{pmatrix}, 
\label{S20}
\end{equation}
in which the complex variable $z$ is expressed as $e^{ik}$ under periodic boundary conditions. The characteristic equation is written as $F(z,E)=\det[H(z)-E I_{2\times 2}]=\frac{P_{m+n}(z)-E^2 z^m}{z^m}=0$, with $m=2$ and $n=4$. According to the arguments above, the GBZ is constructed by $|z_m(E)|=|z_{m+1}(E)|$, which has been confirmed numerically in Fig.~\ref{figs5}. 
	
Form Fig.~\ref{figs5}, one can extract the following information: 
(i:) GBZ is formed by $z_2$ and $z_3$, which are roots of the characteristic equation regarding eigenvalues of $H$, and this ensures that GBZ contains the same number $m$ of zeros and poles. 
(ii:) $E$ and $-E$ always correspond to the same roots due to sublattice symmetry, hence the two-part connected bulk spectrum always circulates itself and encloses zero area as $z$ varies along the GBZ, and their behavior is symmetric about the origin. So one can always map a two-band sublattice model to a single-band Hamiltonian.

\section{The Proof for case-(iii)}\label{apdnx:4}

Here we give a rigorous proof for the case (iii) in the main text. In this case, one part of GBZ outside and another part inside the unit circle(BZ), then we define the region inside GBZ but outside BZ as $U$, the region inside BZ but outside GBZ as $V$, and the region inside both GBZ and BZ as $W$. Before the proof, we must display two facts, one is that GBZ must enclose the origin, and another is that GBZ formed by $H^{-1}_m({\mathcal{L}_{GBZ})}$ and $H^{-1}_{m+1}({\mathcal{L}_{GBZ})}$ always encircles the first $m$ zeros that have ordered by absolute value. 

\par Then we first prove that for any choice of base energy $E_b$, the roots of $H(z)-E_b=0$ can not appear in $U$ and $V$ regions at the same time. An obvious fact is that the magnitude of $z^{\prime}_0$ inside $V$ is always less than that of $z_0$ inside $U$ region. Assuming that there exist at least two zeros $z_0$ and $z^{\prime}_0$ correspond to the same base energy $E_b$, then it will happen that for the $E_b$, the root $z^{\prime}_0$ with smaller absolute value does not belong to GBZ but the root $z_0$ with larger absolute value belongs to. This distinctly contradicts the facts we have displayed. Hence the assumption is not true, that is to say, there are no two roots at most correspond to the same $E_b$, i.e., the roots of $H(z)-E_b=0$ can not appear in $U$ and $V$ regions at the same time for any $E_b$. Furthermore, it comes to the conclusion that the other roots of base energy corresponding to $z^{\prime}_0$ may only appear in the region $W$ but not in the region $U$. 

\par The next steps are the same as in case (i) and (ii). Pick $z_0 \in U$ and $E_0=H(z_0)$, then $z_0$ is a zero of $H(z)-E_0$, likely pick $z^{\prime}_0 \in V$ and $E_0^{\prime}=(z^{\prime}_0)$, then $z^{\prime}_0$ is a zero of $H(z)-E_0$. Here we notice that $E_0 \neq E^{\prime}_0$ that has been proved above. With the fact that GBZ encloses $m$ zeros, therefore in case (iii) there are always $E_0$ and $E_1$ such that $w_{BZ,E_0}<-1$ and $w_{BZ,E^{\prime}_0}>1$, and $E_0$ is not equal to $E^{\prime}_0$. Here the case (iii) has been rigorously proved.


\section{Prove the inverse statement of Eq.(7)}\label{apdnx:5}
In this section, we prove the statement that if there is any $E_b \in \mathcal{C}$ with respect to which $H(z)$ has nonzero winding, then one can always find some $n(H,H^{\ast})\neq 0$ such that $J \neq 0$. The natural extension of the definition of total current from Hermitian systems is that 
\begin{equation}
J=\int_0^{2\pi} n(H,H^{\ast})\frac{dH(k)}{dk}dk=\oint_{\mathcal{L}_{BZ}}n(H,H^*)dH. 
\label{Current1}
\end{equation}
If the interior area of $\mathcal{L}_{BZ}$ is nonzero, it means that $\mathcal{L}_{BZ}$ is composed of one or several close loops. One can always find the base energy $E_b$ surrounded by one closed-loop $\mathcal{L}^{\prime}_{BZ} \subseteq \mathcal{L}_{BZ}$. Here we denote the interior area of $\mathcal{L}^{\prime}_{BZ}$ as $S(\mathcal{L}^{\prime}_{BZ})$, and we have 
\begin{equation}
S(\mathcal{L}^{\prime}_{BZ}) \neq 0
\end{equation}
Hence we can always define the distribution function $n(H,H^*)$ as follows
\begin{equation}
\left\{
\begin{array}{lr}
n(H,H^{\ast})=\frac{1}{2i}(H-H^{\ast}), 
&H \in \mathcal{L}^{\prime}_{BZ} \\
n(H,H^{\ast})=0, 
&H \in \mathcal{L}_{BZ} \setminus \mathcal{L}^{\prime}_{BZ}
\end{array}  
\right.,
\end{equation}
such that 
\begin{equation}
J=\oint_{\mathcal{L}_{BZ}}n(H,H^*)dH 
= \oint_{\mathcal{L}^{\prime}_{BZ}}\Im(H) dH, 
\end{equation}
obviously, the imaginary part of which is zero. Then the total current  becomes
\begin{equation}
J=\oint_{\mathcal{L}^{\prime}_{BZ}}\Im(H) d\Re(H)
=S(\mathcal{L}^{\prime}_{BZ}) \neq 0.
\end{equation}
The finite area enclosed by $\mathcal{L}_{BZ}$ ensures the existence of the current under periodic boundary conditions, and which collapses into skin modes with open boundary. Here the inverse statement of Eq.(7) has been proven. 

\section{The proof related to Eq.~(8)}\label{apdnx:6}
In this section, we mainly discuss the following aspects: (i:) Derive the expression of current that is a linear response of the system to infinitesimal gauge transformation; (ii:) Give a general expression of density operator, which returns to the conventional form in the hermitian case, and describes the steady state in non-hermitian case. (iii:) Prove that the persistent current always vanishes for any hermitian system. 

In general, a non-hermitian tight-binding Hamiltonian in real space can be expressed as 
\begin{equation}\label{Hamiltonian:V}
H_0=\sum_{i,j,a,b}h_{ab}^{i j} |a_i \rangle \langle b_j |
\end{equation} 
with $ i,j $ lattice sites and $ a,b$ as the band indices for multi-band system. If the tight-binding matrix satisfies $ h_{ab}^{ij}=h_{ba}^{ji*} $, then the Hamiltonian reduces to hermitian. Due to the translational symmetry of the lattice, the matrix elements $ h_{ab}^{ij}$ are only related to the hopping distance $ r=i-j $ and written as $ h_{ab}^r $. We now take the Fourier transform, 
$ |a_{i}\rangle=\frac{1}{\sqrt{N}}\sum_k e^{-iki}|a_k\rangle $, $ \langle b_j|=\frac{1}{\sqrt{N}} \sum_k e^{ikj} \langle b_k| $ and $ h_{ab}(k)=\frac{1}{\sqrt{N}}\sum_{r}e^{ikr}h_{ab}^r $, where $ N $ is the number of lattice sites. 

Then we take the gauge transformation in the lattice Hamiltonian, the hopping element has changed from $ h_{ab}^{ij} $ to $ h_{ab}^{ij} e^{i\int_{j}^{i}A_{j+r/2} dl}=h_{ab}^{ij} e^{i A_{j+r/2} r}$. Here we assume that the vector potential is the same on the straight line connected by the two sites $ i $ and $ j $, and labeled as $ A_{j+r/2} $. Since this vector potential is defined on the lattice, it can be taken Fourier transform 
$ A_{j+r/2}=\frac{1}{\sqrt{N}}\sum_{q}e^{-iq(j+r/2)} A_q$ with $ q $ the momentum quantum number.
The Hamiltonian Eq.~(\ref{Hamiltonian:V}) after gauge transformation is expressed as
\begin{equation}\label{GTHam:V}
H=\sum_{j,r,a,b}h_{ab}^r e^{i A_{j+r/2} r}|a_{j+r} \rangle \langle b_j | = \sum_{j,r,a,b}h_{ab}^r (1+i A_{j+r/2} r+o(A))|a_{j+r} \rangle \langle b_j |, 
\end{equation}
where the gauge potential $ A $ is infinitesimal, so the higher-order term can be ignored. We focus on the linear term $ H_{l} \equiv \sum_{j,r,a,b}h_{ab}^r i A_{j+r/2} r |a_{j+r} \rangle \langle b_j | $, which can be transform into momentum space, 
\begin{equation}\label{LinearKSpace:V}
\begin{split}
H_{l}&=\frac{1}{N} \sum_{a,b}\sum_r (i r h_{ab}^r) \sum_{k_1,k_2,q}\delta_{k_1-q,k_2} A_q e^{-i(k_1+q/2)r} |a_{k_1} \rangle \langle b_{k_2}| \\
&=\frac{1}{\sqrt{N}}\sum_{a,b}\sum_{k_1,q}\sum_r A_q (i r h_{ab}^r e^{i(-k_1-q/2)r}) |a_{k_1}\rangle \langle b_{k_1-q} |. 
\end{split}
\end{equation}
According to the Fourier transform formula of $ h_{ab}(k) $, one can obtain that $ dh_{ab}(k)/dk = \frac{1}{\sqrt{N}}\sum_r ire^{ikr}h_{ab}^r $. Replace it into the above equation and let $ k\equiv -k_1-q/2 $, then we get 
\begin{equation}\label{LinearTerm:V}
H_l= \sum_{q} A_q  \sum_{a,b,k} \frac{\partial h_{ab}(k)}{\partial k} |a_{-k-q/2} \rangle \langle b_{-k-3q/2}|  = \sum_{q} A_q J_q
\end{equation}
Hence the persistent current serves as a linear response of system to infinitesimal gauge transformation is shown as
\begin{equation}\label{Current:V}
J_q = \frac{\partial H(k+A_q)}{\partial A_q}|_{A=0} = \sum_{a,b,k}\frac{\partial h_{ab}(k)}{\partial k} |a_{-k-q/2} \rangle \langle b_{-k-3q/2}| 
\end{equation}
with $ q $ the momentum quantum number. 

Here we assume the density operator in non-hermitian system as 
\begin{equation}\label{DensityOpe:V}
\rho=\sum_{n,k}n(E_{n,k},E_{n,k}^*)|n_k^R \rangle \langle n_k^L|, 
\end{equation}
where $ n(H,H^*) $ refers to state distribution function that only depends on the energy of the state but does not depend on k explicitly. And the superscript $ R,L $ denote the right and left eigenvectors respectively. These eigenvectors satisfy orthogonality and normality, 
\begin{equation}\label{Orthonormality:V}
\sum_{n,k}|n_k^R \rangle \langle n_k^L | =1; \quad  \langle n_k^L | m_{k'}^{R} \rangle =\delta_{nm}\delta_{kk'}. 
\end{equation} 
Note that in hermitian case the $ R,L $ superscripts can be removed, then Eq.~(\ref{DensityOpe:V}) returns to the conventional expression. The eigen equations of the left and right vectors in the non-Hermitian system can be expressed as
\begin{equation}\label{Eigenequation}
H |n_k^R \rangle = E_{n,k}  |n_k^R \rangle; \quad
H |n_k^L\rangle =  E_{n,k}^* |n_k^L\rangle. 
\end{equation}
According to these preliminary preparations, the time evolution of density operator is obtained
\begin{equation}\label{TimeEvolution:V}
\frac{\partial \rho(t)}{\partial t} = \sum_{n,k} 
\frac{\partial n(E_{n,k},E_{n,k}^*)}{\partial t}|n_k^R(t) \rangle \langle n_k^L(t)| + n(E_{n,k},E_{n,k}^*) \frac{\partial}{\partial t}(|n_k^R(t) \rangle \langle n_k^L(t)|). 
\end{equation}
Due to the properties of biorthogonal vector, the term $ |n_k^R(t) \rangle \langle n_k^L(t)| $ is unchanged under time evolution, 
\begin{equation}\label{TimeEvolution1:V}
\frac{\partial}{\partial t}(|n_k^R(t) \rangle \langle n_k^L(t)|)
=\frac{\partial}{\partial t} ( e^{-iE_{n k}t}|n_k^R(0) \rangle e^{iE_{n k}t}\langle n_k^L(0)| ) = \frac{\partial}{\partial t}(|n_k^R(0) \rangle \langle n_k^L(0)|) = 0, 
\end{equation}
so the second term of Eq.~(\ref{TimeEvolution:V}) equals to zero. If the state distribution function does not change with time, we can finally get 
\begin{equation}\label{SteadyState:V}
\frac{\partial \rho(t)}{\partial t}=0,
\end{equation}
which means that Eq.~(\ref{DensityOpe:V}) describes the density operator of steady state. Additionally, one can note that Eq.~(\ref{DensityOpe:V}) reduces to the conventional expression in hermitian systems, 
\begin{equation}\label{HermiDenOpe:V}
\rho_H = \sum_{n.k} n(E) |n_k \rangle \langle n_k|, 
\end{equation}
where $ n(E) $ is a state distribution function, such as Boltzmann distribution, Bose distribution, Fermi distribution etc., and eigenvalue $ E $ is real number due to the requirement of hermiticity. 

\begin{figure}
	\begin{centering}
		\includegraphics[width=0.7\linewidth]{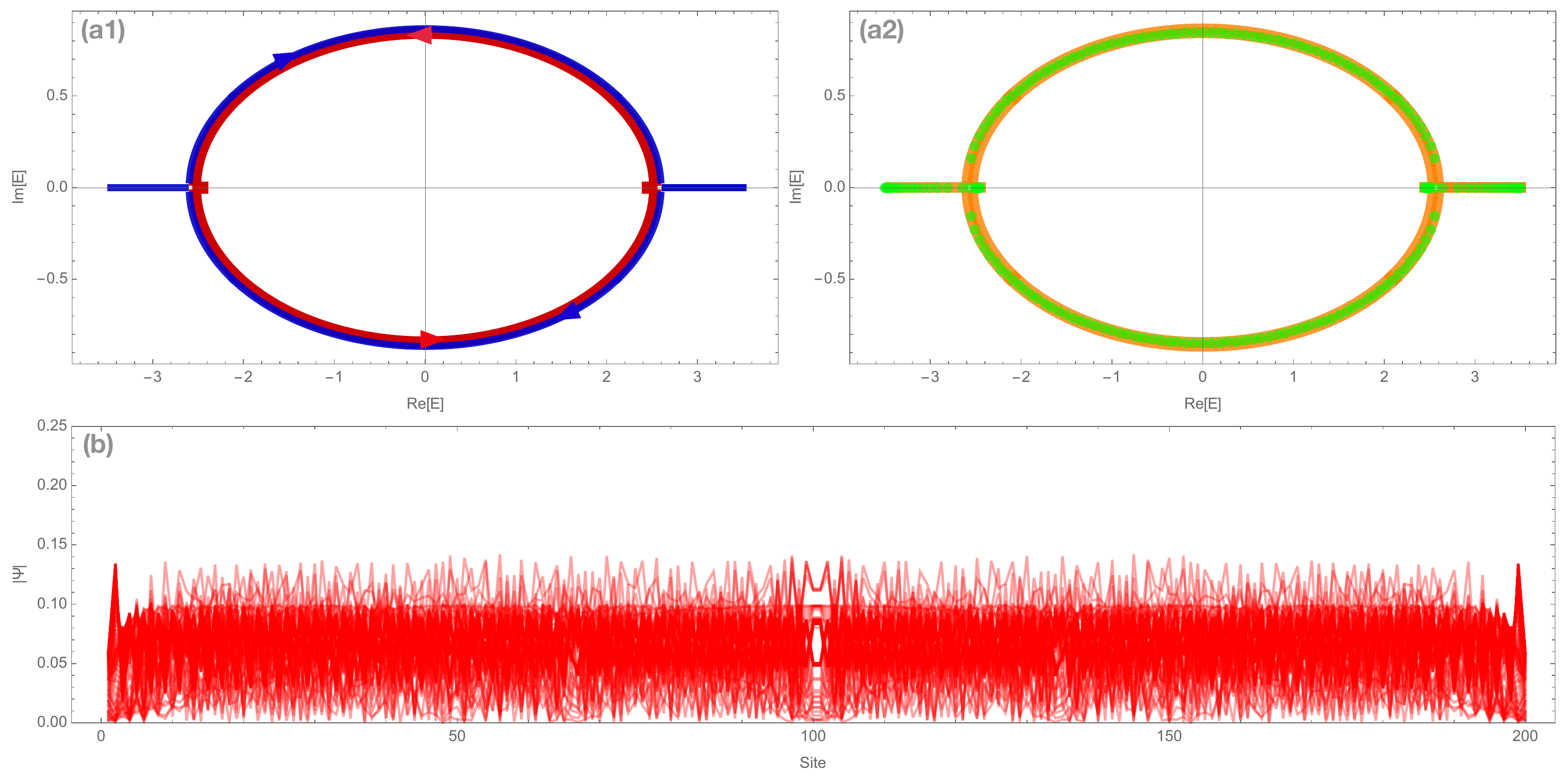}
		\par\end{centering}
	\protect\caption{\label{figs7}The two-band tight-binding model is chosen as $ H(k)=3\cos k \sigma_0 +\frac{1}{2}\sigma_x - i\sin k \sigma_z $. (a1) shows that two bands have opposite orientation as $ k $ varies from $ 0 $ to $ 2\pi $, hence the individual winding number with opposite sign for each band can be defined and total winding vanishes. (a2) illustrates that periodic-boundary spectrum with orange color coincides with open-boundary spectrum with green color. (b) shows the distribution of eigenvectors of the two-band Hamiltonian, which means there is no skin modes. 
	}
\end{figure}

In the end, the general expression of persistent current, according to Eq.~(\ref{Current:V}) and Eq.~(\ref{DensityOpe:V}), can be evaluated as 
\begin{equation}\label{CurrentDerivation:V}
\begin{split}
J&=\mathrm{Tr} (\rho J_q) = \mathrm{Tr}( \sum_{n,k',a,b,k}n(E_{n,k'},E_{n,k'}^*)|n_{k'}^R \rangle \langle n_{k'}^L| \frac{\partial h_{ab}(k)}{\partial k} |a_{-k-q/2} \rangle \langle b_{-k-3q/2}| ) \\
&= \sum_{n,k',a,b,k} n(E_{n,k'},E_{n,k'}^*) \frac{\partial h_{ab}(k)}{\partial k} \mathrm{Tr}( |n_{k'}^R \rangle \langle n_{k'}^L|a_{-k-q/2} \rangle \langle b_{-k-3q/2}| ) \\
&= \sum_{n,k',a,b,k} n(E_{n,k'},E_{n,k'}^*) \frac{\partial h_{ab}(k)}{\partial k}\delta_{n,a}\delta_{n,b}\delta_{k',-k-q/2}
\delta_{k',-k-3q/2}, 
\end{split}
\end{equation}
in which only q = 0 contributes to the current. Then it follows that
\begin{equation}\label{PersistentCurrent:V}
J = \mathrm{Tr} (\rho J_q) = \mathrm{Tr} (\rho J_0) =\sum_{n,k} n(E_n,E_n^*) \frac{\partial E_{n,k}}{\partial k}. 
\end{equation}
The sum of $ k $ is converted into the form of integral in continuum limit, and the above equation can be further expressed as
\begin{equation}\label{PersistentCurrent1:V}
J= \sum_{n}J_n= \sum_{n} \int_{0}^{2\pi} n(E,E^*) \frac{\partial E_{n,k}}{\partial k} dk
=\sum_{n} \oint_{\mathcal{L}_{n,BZ}} n(E_n,E_n^*) dE_{n}. 
\end{equation}
In this equation $ J_n $ represents the current component for $ n_{th} $ band, and equals to zero when the corresponding spectrum of the band, labeled as $ \mathcal{L}_{n,BZ} $, has no interior. 
For a general non-hermitian system, zero area enclosed by energy spectrum is equivalent to the absence of skin modes under open boundary condition. 
While $J_i=0$ implies $J=0$, $J=0$ does not necessitate $J_i=0$ for each $i$. As illustrated in Fig.~\ref{figs7}, these two-band system has individual nonzero current for each band, but zero total current for all bands. 
Therefore, $J=0$ is equivalent to the collapse of the spectrum, not of each single band, but of all bands, into a curve that has no interior.
In hermitian systems, the density operator reduces to Eq.~(\ref{HermiDenOpe:V}), then this persistent current becomes
\begin{equation}\label{HermiPerCur:V}
J_H = \mathrm{Tr}(\rho_H J_0) = \sum_{n} \oint_{\mathcal{L}_{n,BZ}} n(E_n) dE_{n}, 
\end{equation}
where $ n $ is band indices and $ E_n $ is real number due to hermiticity. For each band, the area enclosed by the integral loop is zero since the spectrum of hermitian systems is always ``step-back" arcs, hence the integral Eq.~(\ref{HermiPerCur:V}) is always zero, namely 
\begin{equation*}
J_H=0 
\end{equation*}
Here we completed all the proof in this section, and it comes to the final conclusion that the persistent current is derived as a linear response, and the persistent current vanish for any Hermitian system.


\end{document}